\documentclass[11pt]{amsart}
\usepackage{amsbsy,amsfonts}
\usepackage{amstext}

\usepackage{latexsym,eucal,amsmath,amsthm}
\usepackage{amssymb}
\newtheorem{definition}{Definition}[section]
\newtheorem{theorem}{Theorem}[section]
\newtheorem{lemma}{Lemma}[section]
\newtheorem{corollary}{Corollary}[section]
\newtheorem{proposition}{Proposition}[section]
\newtheorem{remark}{Remark}[section]
\numberwithin{equation}{section}
\begin{document}

\title{Anderson localization  for the discrete one-dimensional quasi-periodic  Schr\"{o}dinger operator with potential defined by a Gevrey-class function}
\author{Silvius Klein}
%\keywords{Discrete quasi-periodic  Schr\"{o}dinger operator, Anderson localization, Lyapunov exponent, Integrated density of states.}
%\subjclass{81Q10, 47B39, 37D25, 82B44}
%\renewcommand{\subjclassname}{\textup{2000} Mathematics Subject Classification}
\date{December, 2003}
\thanks{The author was supported in part by NSF grant DMS-9970469 and by NSF grant DMS-9985572}
\begin{abstract}{In this paper we consider the discrete one-dimensional\\ 
Schr\"{o}dinger operator with quasi-periodic potential 
$v_n = \lambda v (x + n \omega)$. We assume that the frequency $\omega$ 
satisfies a strong Diophantine condition and that the function $v$ belongs to a 
Gevrey class, and it satisfies a transversality  condition. Under these 
assumptions we prove - in the perturbative regime - that for large disorder
 $\lambda$ and for most frequencies $\omega$ the 
operator satisfies Anderson localization. Moreover, we show that the associated Lyapunov exponent is positive for all energies, and that the Lyapunov exponent and the integrated density of states are continuous functions with a certain modulus of continuity.
We also prove a partial nonperturbative result assuming that the function $v$ belongs to some particular Gevrey classes.}
\end{abstract}

\address{Department of Mathematics, UCLA, Los Angeles, CA  \& IMAR, Bucharest}
\email{kleins@math.ucla.edu}

\maketitle

\section{Introduction and statements}

The discrete one-dimensional Schr\"{o}dinger operator with quasi-periodic potential is the selfadjoint, bounded operator $ H(x) = H_{\omega, \lambda} (x)$ on $l_2 (\mathbb{Z})$ defined by
\begin{equation}\label{op1} 
H_{\omega, \lambda} (x) :=  - \Delta +  \lambda v( x + n \omega ) \delta_{n,n'}
\end{equation} 
where $\Delta$  is the discrete (lattice) Laplacian on $l_2 (\mathbb{Z})$  : 
\begin{equation}\label{laplace} 
(\Delta u)_n  :=  u_{n+1} +  u_n
 \end{equation} 

In (\ref{op1}), $v $ is a real valued function on $ \mathbb{T} = \mathbb{R} / 2 \pi \mathbb{Z} $, that is, a real valued 
$ 2\pi$-periodic function on $\mathbb{R} $, $x$ is a parameter on $ \mathbb{T} $, $\omega $ is an irrational frequency and  $\lambda$ is a real number called the disorder of the system. 

\medskip

We may assume the following on the data : 

\begin{itemize}
 \item (Strong) Diophantine condition on the frequency :
$ \omega \in DC_{\kappa} \subset \mathbb{T} $ for some $ \kappa > 0 $. 
That is, 
\begin{equation}\label{DC}
\mbox{dist } (k \omega , 2 \pi \mathbb{Z} )  =:  || k \omega || > \kappa \cdot 
\frac{1} {| k |(\log(1 + | k |))^{3}} \hspace{.2in} \forall \, k \in  \mathbb{Z} \backslash \{ 0 \} 
\end{equation} 
Notice that $ \mbox{mes } [ \mathbb{T} \backslash DC_{\kappa} ] \lesssim \kappa $. 
\end{itemize}

\begin{itemize}
 \item Gevrey-class regularity on the function: $v$ is a smooth function which belongs to a Gevrey class $G^{s} (\mathbb{T}) $ for some  $ s > 1 $. 
That is,
\begin{equation}\label{GC}
\sup_{ x \in \mathbb{T}}  | \partial ^{m} v(x)| \leq M  K^{m} ( m ! )^{s}  \hspace{.2in} 
\forall \, m \geq 0 
\end{equation}
for some constants $M,$ $K$ $ > 0 $.

This condition is equivalent (see Chapter V.2 in \cite{Ka}) to the following exponential-type decay of the Fourier coefficients of $v$:
\begin{equation}\label{fcoef} 
| \hat{v} (k) | \leq  M e^{- \rho |k|^{1/s}} \hspace{.2in} \forall \, k \in \mathbb{Z}
\end{equation} 
for some constants $M,$ $\rho$ $ > 0 $ , where 
\begin{equation}\label{fexp}
v(x) = \sum_{k \in \mathbb{Z}} \hat{v} (k) e^{i k x}
\end{equation} 
We will use (\ref{fcoef}) instead of (\ref{GC}).
\end{itemize}

\begin{itemize}
 \item Transversality condition on the function : $v$ is not flat at any point. That is :
 \begin{equation}\label{TC}
\forall \, x \in \mathbb{T} \hspace{.1in}  \exists \, m \geq 1 \hspace{.1in} \mbox{ so that }  \partial ^{m} v(x) \neq 0
\end{equation}
\end{itemize}

\medskip

Notice from (\ref{GC}) or (\ref{fcoef}) with $s = 1$ that the Gevrey class $ G^{1} (\mathbb{T})$ is the class of analytic functions on $\mathbb{T} $. The transversality condition (\ref{TC}) on a 
function in this class, simply means that the function is non constant. Therefore, the Schr\"{o}dinger operator with a potential given by a function which satisfies the Gevrey class regularity condition (\ref{GC}) and the transversality condition (\ref{TC}) is a natural generalization of the non constant analytic case considered in \cite{B}, \cite{BG}, \cite{GS}.
 
Notice also that $ s_1  <  s_2 \hspace{.1in} \Rightarrow \hspace{.1in} G^{s_1} (\mathbb{T}) \subset G^{s_2} (\mathbb{T}) $, so the greater the order of the Gevrey class is, the larger the class becomes. 

Let's recall the following definitions (see also \cite{B}):

\begin{definition}\label{ALdef}
\rm{We say that an operator satisfies Anderson localization if it has pure point spectrum with exponentially decaying eigenfunctions.
}
\end{definition}

\begin{definition}\label{lyapdef}\rm{
Consider the  Schr\"{o}dinger equation:
$$H_{\omega, \lambda} (x) u  =  E u $$
for $u = [ u (n) ]_{n \in \mathbb{Z}} \subset  \mathbb{R}$. Then

$$
\left( \begin{array}{cc}
u (n +1)\\
u(n) \\  \end{array} \right)  = M_{N} (E)  \left( \begin{array}{cc}
u (1)\\
u(0) \\  \end{array} \right)$$
where $$ M_{N} (E) = M_{N} (x, \omega, \lambda, E) :=
\prod_{j=N}^{1}  \left( \begin{array}{ccc}
\lambda v (x + j \omega ) - E  & &  - 1  \\
1 & &  0 \\  \end{array} \right) $$ 
is called the transfer (or fundamental) matrix of  (\ref{op1}).

Define further $$L_{N} (E) = L_N (E, \omega, \lambda) := \int \frac{1}{N} \log || M_N (x, E) || \, dx$$ 
and $$L(E) := \lim_{N \rightarrow \infty} L_N (E) $$

$L (E) $ is called the Lyapunov exponent of  (\ref{op1}).
}
\end{definition}

\begin{definition}\label{idsdef}\rm{
For any interval  $\Lambda \subset \mathbb{Z}$ centered at the origin, let $E_{\Lambda} (x) $ denote the set of eigenvalues of the operator $H (x)$ restricted to $\Lambda$ with Dirichlet boundary conditions. Set 
$$ N_{\Lambda} (E, x) := \frac{1}{| \Lambda |} \, \# [ (- \infty, E) \cap E_{\Lambda} (x) ] $$
The ergodic theorem implies that the (weak) limit (in the sense of measures)
$$\lim_{| \Lambda | \rightarrow \infty} N_{\Lambda} (E, x) =: N (E) = N_{\omega, \lambda} (E)  $$
exists for a.e. $x \in \mathbb{T}$ (and it does not depend on $x$ up to a set of measure $0$).

$N (E) $ is called the integrated density of states (IDS) of the operator $ H(x) $ and it is linked to the Lyapunov exponent by the Thouless formula:
$$ L (E) = \int \log | E - E' | \, d N (E') $$
}
\end{definition}

\smallskip

The main result of this paper is the following : 

\begin{theorem}\label{main}
Consider the Schr\"{o}dinger operator (\ref{op1}) 
$$ H_{\omega, \lambda} (x) :=  - \Delta +  \lambda v( x + n \omega ) \delta_{n,n'} $$
Assume that $ v \in G^{s} (\mathbb{T})$, where $s > 1 $, $v$ satisfies the transversality condition (\ref{TC}) and $ \omega \in DC_{\kappa} $  for some $ \kappa > 0 $. There exists $\lambda_{0} = \lambda_{0} (v, \kappa)$ so that the following hold:
 
\mbox{\rm{(P)}} For $ | \lambda | \geq \lambda_{0} $, the Lyapunov exponent of (\ref{op1}) is positive for all energies $E \in \mathbb{R} $:
\begin{equation}\label{lyap1} 
L_{\omega, \lambda} (E) \geq \frac{1}{4} \log | \lambda |  > 0 
\end{equation}

\mbox{\rm{(C)}} For $ | \lambda | \geq \lambda_{0} $, the Lyapunov exponent $L_{\omega, \lambda} (E)$ and the integrated density of states  $N_{\omega, \lambda} (E)$ are continuous functions of the energy $E$, with modulus of continuity - on any compact interval $I$ - at least as good as 
\begin{equation}\label{modcont}
h (t) = C \, e^{- c  |\log t |^{\eta}}
\end{equation}
where $ C = C( I, \lambda, v, \kappa, s) $ and $c$, $\eta$ are some positive absolute constants.
 
\mbox{\rm{(AL)}} Fix $ x_0 \in \mathbb{T} $, and $\lambda$  so that $ | \lambda | \geq \lambda_{0}$. Then for a.e. frequency $\omega \in DC_{\kappa}$, $H_{\omega, \lambda} (x_0) $ satisfies Anderson localization. 
\end{theorem}

The theorem above is a perturbative (so weaker) result, because the size of the disorder $\lambda$  depends on $ \kappa $ too (not only on $v$), so there is a a dependence  on the 
frequency $\omega$. However, when $v$ is 'close' to being analytic, we can prove the following:
\begin{theorem}\label{sec}
Consider the Schr\"{o}dinger operator 
\begin{equation}\label{op2}
H_{\omega} (x) :=  - \Delta +  v(x + n \omega ) \delta_{n,n'}
\end{equation} 
Assume that $v \in G^{s} (\mathbb{T})$, where $ 1 < s < 2 $ and $ \omega \in DC_{\kappa} $  for some $ \kappa > 0 $.
Assume also that the Lyapunov exponent of (\ref{op2}) is positive
\begin{equation}\label{posi2}
L_{\omega} (E) \geq c_0 > 0
\end{equation}
for all $ \omega \in DC_{\kappa} $ and for all $E \in  I $, where $I$ is  some compact interval. Then we have:

\mbox{\rm{(C)}} The Lyapunov exponent $L_{\omega} (E)$ and the integrated density of states  $N_{\omega} (E)$ are continuous functions on $I$, with modulus of continuity at least 
\begin{equation}\label{npmodcont}
h (t) = C \, e^{- c  |\log t |^{\eta}}
\end{equation}
where $ C = C (I, v, \kappa, s) $, $c$ is some positive universal constant and \\
$ \eta = \eta  (v, \kappa, s) $, with $\eta \rightarrow 0$ as $s \rightarrow 2$.

\mbox{\rm{(AL)}} Assume that (\ref{posi2}) holds for a.e. $\omega \in DC_{\kappa} $ and for all $E \in \mathbb{R} $. Fix $x_0 \in \mathbb{T} $. Then for a.e. $ \omega \in DC_{\kappa} $, $H_{\omega} (x_0) $ satisfies Anderson localization. 
\end{theorem}
\begin{remark}\label{thm2r} \rm{The disorder in Theorem \ref{sec} is fixed. We don't assume the transversality condition (\ref{TC}) on $v$, but we assume the positivity of the Lyapunov exponent. 

The statement (C) in this theorem extends the continuity result of Theorem 6.1 in \cite{GS} from a potential given by an analytic function to one given by a more general Gevrey-class function of order $ s < 2 $. We are not able to get H\"{o}lder continuity as in \cite{GS} though, but only the weaker modulus of continuity (\ref{npmodcont}).

The statement (AL) in this theorem extends the localization result of Theorem 10.1 in \cite{B} or of Theorem 1 (for dimension 1) in \cite{BG} from a potential given by an analytic function to one given by a more general Gevrey-class function of order $ s < 2 $.

Therefore - at least when $ v \in G^{s} (\mathbb{T}) $ with $s < 2 $ - the problem of showing continuity of the Lyapunov exponent and of the IDS and Anderson localization for the operator (\ref{op2}) is reduced to the one of proving positivity of the Lyapunov exponent.} 
\end{remark}

There is a long list of related results in the literature. We will mention only the ones that are most relevant to this paper : 

\textbf{-} In 1991, E. Sorets and T. Spencer considered (see \cite{S-S}) the operator $H_{\omega, \lambda} (x) $ given by (\ref{op1}) - with any frequency 
$\omega$ and a nonconstant analytic function $v$. They proved that for $ |\lambda|  \geq \lambda_{0} $, where $\lambda_0$ depends only on $v$, the Lyapunov exponent is bounded away from zero for all energies $E$: $L (E) > \frac{1}{2} \log | \lambda| $.

\textbf{-} In 1997, L. H. Eliasson considered (see \cite{E}) the operator $H_{\omega, \lambda} (x) $ given by (\ref{op1})  - with  frequency $\omega$ satisfying a (weak) Diophantine condition and the function $v$ satisfying the Gevrey-class regularity and the transversality condition. Under these assumptions, he proved - using KAM methods - that for  $ |\lambda|  \geq \lambda_{0} $, where $\lambda_0$ depends on the function $v$ and on the Diophantine condition on $\omega$, the operator $ H_{\omega, \lambda} (x)$ has pure point spectrum for a.e. $x \in \mathbb{T}$. Moreover, this implies, using Kotani's theory (see \cite{Simon}, \cite{B}) that the Lyapunov exponent is non zero for 
a.e. energy $E$. The author has also suggested that the argument could be modified to obtain exponential decay of the eigenfunctions, but he has not provided a proof of it.

\textbf{-} In 1999, S. Jitomirskaya proved (see \cite{J}) Anderson localization for the \\ Almost-Mathieu operator, that is, for the operator (\ref{op1}) with $ v(x) = \cos x $. The result in \cite{J} is nonperturbative (and very precise): for any Diophantine $\omega $, for a.e. $x$, and for 
$ | \lambda | > 2 $, there is only point spectrum with exponentially decaying (at exactly the 
Lyapunov rate) eigenfunctions. 

\textbf{-} In 2000, J. Bourgain and M. Goldstein considered (see \cite{BG}, \cite{B}) the operator $ H_{\omega} (x) $ given by (\ref{op2}) - 
where $\omega$ satisfies a weak Diophantine condition and $v$ is a nonconstant analytic function. Assuming also that the Lyapunov exponent is positive: 
$ L_{\omega} (E) > 0 $ for a.e. $\omega$ and for all $E$, the authors prove that the operator  $H_{\omega} (x) $ satisfies  Anderson localization - with exponential decay of the eigenfunctions at almost Lyapunov rate -  for every $x$ and for a.e. $\omega$. This - combined with E. Sorets and T. Spencer result mentioned above - implies nonperturbative Anderson localization for the operator  $H_{\omega, \lambda} (x) $ given by (\ref{op1}) assuming $\omega$ is a Diophantine frequency, $v$ is a non constant analytic function, and $\lambda$ is a large enough disorder - depending only on  $v$.  

\textbf{-} In 2001, M. Goldstein and W. Schlag proved (see \cite{GS}) that on every compact interval on which the Lyapunov exponent is bounded away from zero, both the Lyapunov exponent and the IDS are H\"{o}lder continuous functions of the energy. In this paper, the function $v$ defining the potential of $H_{\omega} (x)$ is  non constant and real analytic, while the frequency $\omega$ satisfies a (strong) Diophantine condition.

\textbf{-} In 2001, J. Bourgain, M. Goldstein, W. Schlag proved (see \cite{BGS}) Anderson localization and positivity of the Lyapunov exponents for the skew-shift model, with potential given by a non constant analytic function. Their result is perturbative - the disorder of the system depends on the frequency.

\vspace{.07in}

Our paper shows basically that the methods in \cite{B}, \cite{BG} and especially in  \cite{GS}, \cite{BGS} are robust enough to allow more general potentials, namely those defined by a  Gevrey-class function which also satisfies the transversality condition. We have not been able, though, to get nonperturbative results, other than a partial one, Theorem \ref{sec}. We will follow closely the ideas in the aforementioned papers. 

We prove a large deviation theorem (LDT) for the transfer matrices associated to (\ref{op1}) or (\ref{op2}). As in \cite{BGS} and \cite{GS}, this LDT will be used to prove the positivity of the Lyapunov exponent in the perturbative case, and the continuity of the Lyapunov exponent and of the IDS in both the perturbative and the nonperturbative cases. This LDT will also imply, as in \cite{B}, \cite{BG}, 'good bounds' on the Green's functions associated to (\ref{op1}) or (\ref{op2}). Because of the exponential-type decay (\ref{fcoef}) of the Fourier coefficients of our function $v$, the same arguments - using semi-algebraic set theory - from \cite{B}, \cite{BG}, will apply similarly to this more general situation, eventually proving Anderson localization for these operators.

The challenge is then to prove the LDT for the transfer matrices $M_N (x)$. The LDT says that $u_N (x) := \frac{1}{N} \log || M_N (x) || $ is close to its mean (integral) denoted by $ <u_N> $ for all $x$ outside a small set (where how 'close' or 'small' will be expressed in terms of the scale $N$). 

In \cite{B}, \cite{BG}, where $v$ is an analytic function, this is proved exploiting the existence of a subharmonic extension $ u_N (z)$ of $u_N (x)$. Using the Riesz representation theorem for subharmonic functions, it is shown that the Fourier coefficients of $u_N (x)$ have the decay : 
\begin{equation}\label{decfcoef} 
|\hat{u}_N (k) | \lesssim \frac{1}{|k|} \hspace{.1in} \mbox{ for all } k \neq 0 
\end{equation}
It is important that the decay (\ref{decfcoef}) is uniform in $N$. Using Fourier expansion, (\ref{decfcoef}) implies, for $x$ outside 
a small set, a good approximation of the mean $ <u_N> $ by averages of shifts 
of $u_N (x)$, provided we consider shifts with a Diophantine frequency (so that their 
orbits are fairly uniformly distributed on $\mathbb{T}$). This, combined with the fact 
that $u_N (x)$ is close to averages of its shifts provided the number of shifts considered
is much smaller then the scale $N$, eventually leads to the proof of the LDT. 

For functions which are not analytic, $u_N (x)$ does not have a subharmonic 
extension. The idea is then to substitute - at each scale $N$ - in $M_N (x)$ and  in $u_N (x)$, 
the potential $v (x) $, by an appropriate truncation $v_N (x)$. The new function $u_N (x)$ 
has a subharmonic extension, but the trade-off is that the decay of its Fourier 
coefficients is not uniform in $N$ as in (\ref{decfcoef}), and we only get :
\begin{equation}\label{dfcoef} 
|\hat{u}_N (k) | \lesssim  N^{\delta} \frac{1}{|k|} \hspace{.1in} \mbox{ for all } k \neq 0 
\end{equation}
where $\delta  > 0 $ is a power which depends on how we define the truncations $v_N (x)$, and on the order $s$ of the Gevrey class the function $v (x) $ belongs to.
 
As long as $\delta < 1 $, we can use (\ref{dfcoef}) in a similar way the uniform decay 
(\ref{decfcoef}) is used in \cite{B}, \cite{BG} and therefore prove the LDT (see Theorem \ref{shifts} and Theorem \ref{LDT<2}). This situation corresponds to functions $v$ 'close' to being analytic, namely $v \in G^{s} (\mathbb{T})$ with $s < 2$. 

But in general, for functions in an arbitrarily large Gevrey class, the 
power $\delta $  in the estimate (\ref{dfcoef}) is  $\, \geq 1 $, and (\ref{dfcoef}) is too weak to prove the LDT with this method. The same kind of technical problem, but for a 
different model - the skew shift - was encountered in \cite{BGS}. There, the authors used 
the avalanche principle from \cite{GS} to better control the size of the transfer matrices. We will follow this approach for the general case (see Section 5). 

The paper is organized as follows: in Section 2 we give the basic definitions and we show rigorously the approximation (truncation) argument explained above. In Section 3 we prove a technical result, on Diophantine shifts 
of subharmonic functions, to be used in the proof of the LDT. In Section 
4 we prove the LDT for the case $s < 2$. In Section 5 we prove the LDT in the 
general case, and the positivity of the Lyapunov exponent - which is the statement (P) in Theorem \ref{main}. Using the LDT from previous sections, in Section 6 we complete the proofs of Theorem \ref{main} and of Theorem \ref{sec} by proving the continuity of the Lyapunov exponent and of the IDS (the statement (C) in these theorems) and Anderson localization for the operators (\ref{op1}) or (\ref{op2}) (the statement (AL) in these theorems).

\section{ Definitions, notations, general setup} 

We assume that the function $v = v(x)$  belongs to the Gevrey class $G^{s} (\mathbb{T})$, where $s > 1$ and $ \mathbb{T} = \mathbb{R} / 2 \pi \mathbb{Z} $. Therefore $v(x)$ is a $ 2 \pi$-periodic function on $\mathbb{R}$ so that 
\begin{equation}\label{Fexp}
v(x) = \sum_{k \in \mathbb{Z}} \hat{v} (k) e^{i k x} 
\end{equation}
\begin{equation}\label{Fcoef}
\mbox{where  } \; \;\; | \hat{v} (k) | \leq  M e^{- \rho |k|^{1/s}} \hspace{.2in} \forall \, k \in \mathbb{Z}
\end{equation}
for some constants $M, \rho > 0 $. 

At every scale $N$ we will substitute - in the formula defining the $N$th transfer matrix $M_N (x) $ - the function $v (x)$ by a trigonometric polynomial $v_N (x) $. This polynomial should be chosen to approximate $v (x) $ within a very small error (this error should be (super)exponentially small in $N$), so that the ``transfer matrix substitute'' would be close to the original transfer matrix. Therefore, the degree 
deg $v_N =: \tilde{N}$ of this polynomial should be very large, namely, based on the rate of decay (\ref{Fcoef}) of the Fourier coefficients of $v$, $\tilde{N}$
should be a power of $N$ (which would depend on the Gevrey class).

The trigonometric polynomial $v_N$ has a holomorphic extension to the whole complex plane, but we should restrict it to a strip of width $\rho_N$ so that this extension would be bounded by a constant depending only on $v$ (uniformly in the scale $N$). It turns out that the width  $\rho_N$ of holomorphicity should be 
 $\rho_N \approx ( \mbox{deg }v_N )^{- 1} \approx \tilde{N}^{-1} \approx N^{- \theta}$, for some power $\theta > 0$.

The fact that the ``substitutes'' $v_N (x) $ have different, smaller and smaller widths of holomorphicity, is what creates additional technical problems (compared to the $v(x)$ analytic function case) and also makes this approach to fail for functions $v(x)$ with slower rate of decay of their Fourier coefficients.

We therefore have to find the optimal ``error vs. degree'' approximations of $v(x)$ by trigonometric polynomials $v_N (x) $. Here is the formal calculation:

For every positive integer $N$, consider the truncation
\begin{equation}\label{trunc}
v_N (x) := \sum_{| k | \leq \tilde{N}} \hat{v} (k) e^{i k x}
\end{equation} 
where $\tilde{N} = \mbox{deg } v_N$ will be determined later.

$v_N (x) $ is an analytic, $2 \pi$-periodic function on $\mathbb{R}$. It can be extended to a holomorphic ($2 \pi$-periodic) function on $\mathbb{C}$ by: 
\begin{equation}\label{holext} 
v_N (z) := \sum_{| k | \leq \tilde{N}} \hat{v} (k) e^{i k z}
\end{equation}

To assure the uniform boundedness in $N$ of $v_N (z)$ we have to restrict 
 $v_N (z)$ to the  strip $ [ | \Im z | < \rho_{N} ] $, where $ \rho_{N} := \frac{\rho}{2} \tilde{N}^{1/s - 1} $. Indeed, if $z = x + iy$, $|y| < \rho_{N}$, then: 
$$| v_N (z) | = | \sum_{| k | \leq \tilde{N}} \hat{v} (k) e^{i k z} | \leq
\sum_{| k | \leq \tilde{N}} | \hat{v} (k) | e^{- k y } \leq
M \sum_{| k | \leq \tilde{N}} e^{- \rho | k |^{1/s}}  e^{| k | | y |} \leq $$
$$\leq 2 M \sum_{k = 0}^{\tilde{N}} e^{- \rho | k |^{1/s} + k | y | } \leq
2 M \sum_{k = 0}^{\tilde{N}} e^{- \frac{\rho}{2} k^{1/s}} \leq
2 M \sum_{k = 0}^{\infty} e^{- \frac{\rho}{2} k^{1/s}} =: B < \infty $$
where $B$ is a constant which depends on $\rho, s, M $, and we have used : \\
$ | y | <  \rho_{N} =  \frac{\rho}{2} \tilde{N}^{1/s - 1} \leq  \frac{\rho}{2} | k |^{1/s - 1} $  for $ | k | \leq \tilde{N}$.

We also have $ | v (x) - v_N (x)| \leq C e^{-c \tilde{N}^{1/s}}  $  for all 
$x \in \mathbb{R}$, where $C, c > 0 $ depend on $ \rho, s$. 

We will need, as already mentioned, (super)exponentially small error, so $\tilde{N}$ should be chosen such that $ e^{-c \tilde{N}^{1/s}} \leq e^{- N^{b}}$ for some $b > 1$.

Therefore $\tilde{N} := N^{b \, s}$ for some $b > 1$ to be fixed later, so the width of holomorphicity of $v_N (z)$ will be: $\rho_N  = \frac{\rho}{2} N^{b  s  (\frac{1}{s} - 1)} =  \frac{\rho}{2}  N^{-b (s - 1)}
=: \frac{\rho}{2} N^{- \delta}$, where $\delta := b \, (s - 1) > 0$. 

We conclude: for every integer $N \geq 1$, we have a function $v_N (x)$ on $\mathbb{T}$ so that
\begin{equation}\label{aproxtrunc} 
| v (x) - v_N (x)| < e^{- c N^b} 
\end{equation}
and $v_N (x)$ has a $2 \pi$-periodic holomorphic extension $v_N (z)$ to the strip \\ $ [ | \Im z | < \rho_{N}  = \frac{\rho}{2} N^{- \delta} ] $,   for which 
\begin{equation}\label{boundv} 
|v_N (z)| \leq B
\end{equation} 
where the positive constants $c$, $B$ depend only on $v$. 
The constants $b$, $\delta$ are linked by  $\delta := b (s - 1) \Leftrightarrow b = \frac{\delta}{s - 1} $ and have to satisfy $ b > 1 $ and $\delta > 0$ (so $ s < \delta + 1$).

\medskip

For $v \in G^{s} (\mathbb{T})$, $ \omega \in DC_{\kappa} $, $ \lambda \in \mathbb{R} $ consider the discrete quasiperiodic Schr\"{o}dinger operator (\ref{op1}) :
$$  H(x) = H_{\omega, \lambda} (x)  :=  - \Delta +  \lambda v( x + n \omega ) \delta_{n,n'} $$

Let $ T = T_{\omega} : \mathbb{T} \rightarrow \mathbb{T} $, $ T x := x + \omega $ be the shift by the frequency $\omega $, and let $ T^{j} x = x + j  \omega $ be its $j$th iteration.

For every integer $ N \geq 1 $ and for every energy $E \in \mathbb{R}$, the $N$th transfer matrix  of $H (x) $ is $ M_{N} (x)  =  M_{N} (x, \omega, \lambda, E) := \prod_{j=N}^{1} A (T^{j} x,  \lambda, E) $, where

$$ A(y) = A (y, \lambda,  E) 
 :=  \left( \begin{array}{ccc}
\lambda v (y) - E  & &   - 1  \\
1 & &   0 \\  \end{array} \right)$$  

Denote by  $ L_N (E) =  L_N (E, \omega, \lambda) := \int_{\mathbb{T}} \frac{1}{N} \log || M_N (x, \omega, \lambda, E) || \, dx $.

Then for every energy $E$,
\begin{equation}\label{deflyap}
 L (E) = L_{\omega, \lambda} (E) := \lim_{N \rightarrow \infty} L_N (E, \omega, \lambda) = 
\inf_{N} L_N (E, \omega, \lambda)
\end{equation} 
is the Lyapunov exponent of (\ref{op1}) (see also \cite{B}).
 
We now substitute $v_N (x)$ for $v(x)$ in the definition of the transfer matrix $M_N (x)$ and get : 
$$\tilde{M} _{N} (x)  := \prod_{j=N}^{1} \tilde{A}^{N} (T^{j} x) \hspace{.15in}  \mbox{  where  } \hspace{.1in} \tilde{A}^{N} (y) :=  \left( \begin{array}{cc}\lambda v_{N} (y) - E  &   - 1  \\
1 &   0 \\  \end{array} \right)$$

Denote also by  $\tilde{L}_N (E) = \tilde{L}_N (E, \omega, \lambda) := \int_{\mathbb{T}} \frac{1}{N} \log || \tilde{M}_N (x, \omega, \lambda, E) || \, dx $.

\smallskip

The spectrum of $ H_{\omega, \lambda} (x) $ is contained in the interval $ [ -2 - | \lambda | \, B,  2 + | \lambda | \, B ] $, since $ \sup_{x \in \mathbb{T}} | v (x) | \leq B $. It is then enough to consider only the energies $E$ such that $ |E| \leq 2 + |\lambda| \, B $.

By Trotter's formula,
$$ M_{N} (x) -  \tilde{M}_{N} (x)  =  $$ 
$$ = \sum_{j=1}^{N} A (T^N x) \ldots A (T^{j+1} x) \,  [A (T^j x) - \tilde{A}^{N}(T^j x)] \,  \tilde{A}^{N}(T^{j-1} x) \ldots \tilde{A}^{N} (T x)$$
$$ A (T^j x) - \tilde{A}^{N}(T^j x) = \left( \begin{array}{cc}
\lambda v (T^j x ) - \lambda v_{N} (T^j x )  &   0  \\
0 &   0 \\  \end{array} \right)$$ so 
$$ || A (T^j x) - \tilde{A}^{N}(T^j x) || \leq | \lambda | \, \sup_{y \in \mathbb{T}} \left| v (y) - v_N (y) \right|  <   | \lambda | \, e^{-c N^{b}} $$ 

Moreover : 
$$ || A (T^j x) ||  = ||  \left( \begin{array}{cc}
\lambda v (T^j x ) - E  & - 1  \\
1 & 0 \\  \end{array} \right) ||  \leq  | \lambda | \, B  + |E| + 2  \leq  2 | \lambda | B + 2  \leq  e^{S(\lambda)} $$
so
\begin{equation}\label{boundA}
 || A (T^j x) ||   \leq  e^{S(\lambda)}
\end{equation}
where $ S(\lambda)$ is a (fixed, for fixed $\lambda$) scaling factor : $ 1 \leq  S(\lambda) \approx \log (| \lambda | + C_v )$,
$C_v$ being a constant which depends on $v$. 

Clearly, we also have : 
$$ || \tilde{A}^{N} (T^j x) ||  \leq ||  \left( \begin{array}{cc}
\lambda v_{N} (T^j x ) - E  & - 1  \\
1 & 0 \\  \end{array} \right) ||  \leq | \lambda | \, B + | E | + 2  \leq e^{S(\lambda)} $$  

Therefore, 
$$ ||  M_{N} (x) - \tilde{M}_{N} (x)  || \leq  \sum_{j=1}^{N} e^{S(\lambda)} 
\ldots  e^{S(\lambda)} e^{-c N^b}  e^{S(\lambda)} \ldots  e^{S(\lambda)} \leq  e^{N S(\lambda) - c N^b} $$
and since $b > 1$, if $ N \gtrsim S(\lambda)^{\frac{1}{b - 1}} $, we get
$$ ||  M_{N} (x) - \tilde{M}_{N} (x)  || \leq  e^{-c N^b}
$$

Since $\det M_N (x) = 1 $ and  $\det \tilde{M}_N (x) = 1 $, we have that 
$ \left\| M_N (x) \right\| \geq  1 $ and $ || \tilde{M}_N (x) || \geq  1 $. Thus, for all $ N \gtrsim S(\lambda)^{\frac{1}{b - 1}}$ and for every $x$ 
$$ \left| \frac{1}{N} \log \left\| M_{N} (x) \right\|  -  \frac{1}{N} \log || \tilde{M}_{N} (x) || \, \right|  \leq  \frac{1}{N} || M_{N} (x)  - \tilde{M} _{N} (x) || < e^{-c N^b} $$
and by averaging, 
$$ \left| L_N (E) - \tilde{L}_N (E) \right| < e^{-c N^b}$$

For fixed parameters $ \omega, \lambda, E $ consider: 
$$ u_{N} (x)  := \frac{1}{N} \log || \tilde{M}_{N} (x) || $$ 
and 
$$ < u_{N} >  =  \tilde{L}_{N} := \int_{\mathbb{T}}  u_{N} (x)  dx $$

Since $$ \tilde{M}_{N} (x) = \prod_{j=N}^{1}  \left( \begin{array}{cc}
\lambda v_{N} (x + j \omega ) - E  &  - 1  \\
1 &  0 \\  \end{array} \right) $$
and since $v_N (z) $ is the holomorphic extension of $v_N (x) $ to $ [ | \Im z | < \rho_{N} ] $, it follows that  
$$ \tilde{M}_{N} (z) := \prod_{j=N}^{1}  \left( \begin{array}{cc}
\lambda v_{N} (z + j \omega ) - E  &  - 1  \\
1 &  0 \\  \end{array} \right) $$ 
is the holomorphic extension of $\tilde{M}_N (x) $ to the strip $ [ \left| \Im z \right| < \rho_{N} ] $. 
Using (\ref{boundv}) we get $ \left\| M_N (z) \right\| \leq  S(\lambda)^{N} $. 
Therefore 
$$ u_{N} (z)  := \frac{1}{N} \log || \tilde{M}_{N} (z) || $$ 
is a subharmonic function on the strip $ [ \left| \Im z \right| < \rho_{N} \approx N^{- \delta}] $  so that
for any $z$ in this strip, $ \left| u_N (z) \right| \leq  S(\lambda) $. 

We can summarize all of the above in the following :
\begin{remark} \rm{
For fixed parameters $\omega, \lambda, E$, at every scale $N$, we have a 
$2 \pi$-periodic function $ u_{N} (x)  := \frac{1}{N} \log || \tilde{M}_{N} (x) || $, which extends on the strip  $ [ \left| \Im z \right| < \rho_{N} ] $, 
$ \rho_{N}  \approx  N^{- \delta} $, to a subharmonic function $ u_N (z)$ 
so that 
\begin{equation}\label{boundu}
 \left| u_N (z) \right| \leq  S(\lambda) \hspace{.2in} \forall \, 
z \in  [ | \Im z | < \rho_{N} ]
\end{equation}
(Note that the bound (\ref{boundu}) is uniform in $N$). 

Moreover, if $\delta$ is chosen so that $ s < 1 + \delta $ and if 
$ N \gtrsim  S(\lambda)^{\frac{1}{b - 1}}$, where $ b = \frac{\delta}{s - 1} $ $ > 1$, then the transfer matrices $ M_N (x) $ are well approximated by their substitutes $ \tilde{M}_N (x) $: 

\begin{equation}\label{aproxu}  
\left| \frac{1}{N} \log \left\| M_{N} (x) \right\| -  u_N (x) \right| < e^{-c N^b}
\end{equation}

\begin{equation}\label{aprox<u>} 
\left| L_N -  < u_N >  \right| < e^{-c N^b}
\end{equation}
}
\end{remark}

We will use estimates on subharmonic functions as in \cite{B}, \cite{BG}, \cite{BGS} for the functions $u_N$ in the remark above.
 
The following will be used later :
 
\begin{remark} \rm{
For all $ x \in \mathbb{T} $, and for all parameters $\omega, \lambda, E $,
 we have : 
\begin{equation}\label{mshift} 
 \left| \, \frac{1}{N} \log \left\| M_{N} (x) \right\| -  \frac{1}{N} \log \left\| M_{N} (x + \omega) \right\| \, \right| \leq  \frac{C S(\lambda)}{N}
\end{equation}
where $C$ is a universal constant. 
}
\end{remark}

\begin{proof}
$$  \left| \, \frac{1}{N} \log \left\| M_{N} (x) \right\| -  \frac{1}{N} \log \left\| M_{N} (x + \omega) \right\| \, \right|  = \left| \frac{1}{N} \log \frac{\left\| M_{N} (x) \right\|}{\left\| M_{N} (x + \omega) \right\|}  \, \right| = $$
$$ = \left| \frac{1}{N} \log 
\frac{|| A (T^N x ) \cdot \ldots \cdot A (T^2 x ) \cdot  A (T x ) || }
{|| A ( T^{N + 1} x ) \cdot  A (T^N x ) \cdot \ldots \cdot A (T^2 x ) || } \, \right| \leq $$
$$ \leq \frac{1}{N} \log 
 || ( A(T^{N + 1} x ) )^{- 1} || \cdot  || A (T x ) ||
  \lesssim \frac{S (\lambda)}{N} $$ after using (\ref{boundA}). The inequality (\ref{mshift}) then follows.
\end{proof}

\section{ Averages of shifts of subharmonic functions }

Let $ u = u (x) $ be a function on $\mathbb{T}$ having a subharmonic extension, and $ \omega \in DC_{\kappa} $  for some $ \kappa > 0 $.  We prove that for 
$x$ outside a small set, the mean of $u$ is close to the averages of shifts 
of $u (x)$  by $ \omega$. Here being 'close' 
or 'small' is expressed in terms of the number of shifts considered. To prove stronger 
estimates (see Theorem \ref{shifts}), we have to consider higher order 
averages. In particular (see Corollary \ref{cshifts}) we also get an estimate for first order averages, which is already contained (although not explicitly
formulated) in \cite{B}, \cite{BG}.
 
After writing this paper we have learned that Theorem \ref{shifts} has been proved - even considering only first order averages - in \cite{GS} (see Theorem 3.8 in \cite{GS}). However, we choose to present here our proof, since it gives a different argument - namely an optimization of the one in \cite{B}, \cite{BG}.

Consider the (Fej\'{e}r) kernel (of order $p$) : 
\begin{equation}\label{kernele}
K_{R}^{p} (t) := (\frac{1}{R} \,  \sum_{j=0}^{R-1} e^{i j t}  ) ^{p}
\end{equation}
Then we have :
$$ \left| K_{R}^{p} (t) \right| = \frac{1}{R^p} \left| \frac{1 - e^{i R t}}{1- e^{i t}} \right| ^{p} \leq \frac{1}{R^{p} \left\|t\right\|^{p}} $$
and also $ \left| K_{R}^{p} (t) \right| \leq 1 $  so
\begin{equation}\label{kernelb} 
\left| K_{R}^{p} (t) \right| \leq \min \{ 1, \frac{1}{ R^{p} \left\|t\right\|^{p} } \} \leq \frac{2}{1 + R^{p} \left\|t\right\|^{p} }
\end{equation}
We can write 
\begin{equation}\label{kernelf}
K_{R}^{p} (t) = \frac{1}{R^p} \sum_{j=0}^{p(R-1)} c_{R}^{p} (j) e^{ i j t}
\end{equation}
where $  c_{R}^{p} (j)$  are positive integers so that 
 $$\frac{1}{R^p} \sum_{j=0}^{p(R-1)} c_{R}^{p} (j) = 1$$

Notice that if $p = 1$ then $K_{R}^{1} (t) = \frac{1}{R} \sum_{j = 0}^{R - 1} 
e^{i j t} $ so $ c_{R}^{1} (j) = 1 $ for all $j$. 

\begin{theorem}\label{shifts}
Let $ u : \mathbb{T} \rightarrow \mathbb{R} $, $ \omega \in DC_{\kappa} $  and $\rho > 0 $. Assume that $u (x) $ has a subharmonic extension to the strip 
$ [ | \Im z | < \rho ] $  so that 
\begin{equation}\label{Boundu}
 | u (z) | \leq  S \hspace{.2in} \forall \, z \in   [ | \Im z | < \rho ]
\end{equation} 
Then, if  $ 0 < a < 1 $, $ 0 < \sigma  <  1 - a $ and $ p \in \mathbb{N}$,
$ p > \frac{a}{1 - a} $ we have : 
\begin{equation}\label{shiftldt} 
\mbox{mes }[ x\in \mathbb{T} : | \, \frac{1}{R^p} \sum_{j = 0}^{p ( R - 1)} c_{R}^{p} (j) \cdot u (x + j \omega) - < u > \, | > \frac{S}{\rho} R^{- a} ] < 
e^{- R^{\sigma}}
\end{equation}
for $ R \geq R_0 = R_0 (\kappa, a, \sigma, p) $.
\end{theorem}

\begin{proof}
Fix the numbers $a, \sigma, p $ subject to the constraints in the theorem. 
We may now suppress $p$ from the notations (e.g. $K_R = K_{R}^{p}$,
$ c_R = c_{R}^{p} $).

Choose $ \alpha \in  (a , 1)$ so that $ p > \frac{a}{1 - \alpha} $ 
(which is possible since $ p >  \frac{a}{1 - a} $ ).
 
Since $ u (z) $ is subharmonic on $ [ | \Im z | < \rho ] $  so that 
(\ref{Boundu}) holds, from Corollary 4.7 in \cite{B} we get :
\begin{equation}\label{Riesz} 
| \hat{u} (k) | \lesssim  \frac{S}{\rho} \frac{1}{|k|}
\end{equation}

Expand $u$ as a Fourier series : 
$$ u (x) = < u > + \sum_{k \neq 0} \hat{u} (k) e^{i k x} \hspace{.2in} \forall 
x \in \mathbb{T} $$
$$ \Rightarrow  u (x + j \omega ) = < u > + \sum_{k \neq 0} \hat{u} (k) 
e^{i k (x + j \omega)} \hspace{.2in}  $$
$$ \Rightarrow \frac{1}{R^p} \sum_{j = 0}^{p ( R - 1)} c_{R} (j) \cdot u (x + j \omega) = <u> + \sum_{k \neq 0} \hat{u} (k) \cdot ( \frac{1}{R^p} \sum_{j = 0}^{p ( R - 1)} c_{R} (j) e^{i j k \omega} ) \cdot e^{i k x}$$

Therefore,
\begin{equation}\label{sum1} 
\frac{1}{R^p} \sum_{j = 0}^{p ( R - 1)} c_{R} (j) \cdot u (x + j \omega) - <u>
 =  \sum_{k \neq 0} \hat{u} (k) \cdot K_{R} (k \omega)  \cdot e^{i k x}
\end{equation}
We will estimate the right hand side of (\ref{sum1}).

Since $ \omega \in DC_{\kappa} $, there is a best approximation $ \frac{m}{q} $\, of $\omega$  so that
\begin{equation}\label{besta1}
 R < q < \frac{1}{\kappa} R (\log(1 + R))^{3}
\end{equation} 
thus 
\begin{equation}\label{besta2}
|| j \omega || > \frac{1}{2 q}  \hspace{.2in} \mbox{if } 1 \leq j < q
\end{equation}
(See chapter I in \cite{Lang}).

Split the right hand side of the sum in (\ref{sum1}) as :
\begin{equation}\label{sum2}
\sum_{k \neq 0} \hat{u} (k) \cdot K_{R} (k \omega)  \cdot e^{i k x} = 
   \sum_{ 0 < |k| < R^{\alpha} } \, + \sum_{ R^{\alpha } \leq |k| < q } \, +
 \sum_{|q| \leq |k| < K } \, +  \sum_{|k| \geq K } 
\end{equation}
where $ K = e^{R^{\sigma '}}$ with $\sigma ' \in  (\sigma , 1 - a)$.\\ 
The first three sums in (\ref{sum2}), denoted by (I), (II), (III) 
respectively, will be uniformly bounded in $x$ by $\frac{S}{\rho} R^{-a} $,
 while the forth sum, denoted by (IV) will be estimated in the $L^2$-norm. 
$$ | \mbox{(I)} | \leq \sum_{ 0 < |k| < R^{\alpha} } 
| \hat{u} (k) | \cdot | K_{R} (k \omega) | \leq 
\frac{S}{\rho} \sum_{ 0 < |k| < R^{\alpha} } \frac{1}{| k |} \frac{1}{R^{p} || k \omega ||^{p}}$$
But $ \omega \in DC_{\kappa} $ so $ || k \omega || > \kappa \cdot 
\frac{1} {| k |(\log(1 + | k |))^{3}} $ 
$ \Rightarrow  \frac{1}{| k |  || k \omega ||^{p}} < 
\frac{1}{\kappa ^{p}} | k |^{p - 1} (\log | k | )^{3 p} $

Then
$$  | \mbox{(I)} | \leq \frac{S}{\rho} \frac{1}{\kappa ^{p}} \frac{1}{R^p} R^{\alpha ( p - 1)} (\log R)^{3 p} R^{\alpha} 
=  \frac{S}{\rho}  \frac{1}{\kappa ^{p}}  R^{p (\alpha - 1)} (\log R)^{p} < \frac{S}{\rho} R^{- a}$$
since $p > \frac{a}{1 - \alpha} $ and provided $R$ is large enough, $ R \geq R_o (\kappa, a, p). $

To estimate (II) and (III) we need the following: let $ I \subset \mathbb{Z} $
be an interval of size $|I| < q $. Then for any $k, k' \in I $, 
since $ | k - k' | \leq |I| < q $, (\ref{besta2}) implies
$ || k \omega  - k' \omega ||  >  \frac{1}{2q} $. Arranging the points 
$k \omega $, $k \in I$  according to their distances on the torus to 1,
we get : 
$$ \sum_{k \in I} | K_{R} (k \omega) | 
\leq \sum_{k \in I} \min \{ 1, \frac{1}{R^{p} || k \omega ||^{p}} \} \lesssim
1 + \sum_{1 \leq | j | \leq q} \frac{1}{1 + R^{p} (\frac{j}{q})^{p}} \lesssim
1 + \frac{q}{R} \lesssim \frac{q}{R}$$
Then, for any interval $ I \subset \mathbb{Z} $ of size $ < q$,
\begin{equation}\label{sumI}
\sum_{k \in I} | K_{R} (k \omega) |  \lesssim  \frac{q}{R}
\end{equation}

It follows that: 
$$ | \mbox{(II)} | \leq \sum_{ R^{\alpha } \leq |k| < q } | \hat{u} (k) | \cdot | K_{R} (k \omega) | 
\leq \frac{S}{\rho} \sum_{ R^{\alpha } \leq |k| < q }  \frac{1}{| k |} | K_{R} (k \omega) | \leq $$
$$ \leq  \frac{S}{\rho} \frac{1}{R^{\alpha}}  \sum_{1 \leq |k| < q } | K_{R} (k \omega) | \leq \frac{S}{\rho} \frac{1}{R^{\alpha}}  \frac{q}{R} <
 \frac{S}{\rho} R^{- \alpha} \frac{1}{\kappa} \frac{R (\log (1 + R ))^{3}}{R} <\frac{S}{\rho} R^{- a}$$ 
since $\alpha  > a $, and provided $R \geq  R_0 (\kappa, a)$.
 
Similarly: 
$$ | \mbox{(III)} | \leq \sum_{|q| \leq |k| < K }  | \hat{u} (k) | \cdot | K_{R} (k \omega) | 
\leq \frac{S}{\rho} \sum_{|q| \leq |k| < K }  \frac{1}{| k |} | K_{R} (k \omega) | = $$
$$ =  \frac{S}{\rho} \sum_{1 \leq s \leq K/q} \hspace{.05in} 
\sum_{(s - 1) q \leq | k | < s q}\,  \frac{1}{| k |} | K_{R} (k \omega) | <
\frac{S}{\rho} \sum_{1 \leq s \leq K/q} \frac{1}{s q}  \frac{q}{R} = $$
$$ = \frac{S}{\rho} \frac{1}{R}  \sum_{1 \leq s \leq K/q}  \frac{1}{s }
\approx \frac{S}{\rho} \frac{1}{R} \log \frac{K}{q} 
\leq  \frac{S}{\rho} R^{- 1} R^{\sigma '} < \frac{S}{\rho} R^{- a} $$
since $ \sigma ' < 1 - a $.

We conclude that :
\begin{equation}\label{sum123} 
| \mbox{(I)} |  + | \mbox{(II)} | +  | \mbox{(III)} | < \frac{S}{\rho} R^{-a} 
\end{equation}
uniformly in $ x \in \mathbb{T} $, and for $ R \geq  R_0 (\kappa, a, p)$.
 
We know estimate (IV) in the $L^2$-norm: 
$$ \int_{\mathbb{T}} | \mbox{(IV)} |^{2} = \int_{\mathbb{T}} | \sum_{|k| \geq K} \hat{u} (k) K_{R} (k \omega) e^ {i k x } |^{2} dx =  
\sum_{|k| \geq K} | \hat{u} (k) |^{2} |  K_{R} (k \omega)|^{2} \leq $$
$$ \leq \, \sum_{|k| \geq K} \, |\hat{u} (k) |^{2} < 
(\frac{S}{\rho})^{2} \sum_{|k| > K}  \frac{1}{| k |^2} 
\approx (\frac{S}{\rho})^{2} \frac{1}{K} =
 (\frac{S}{\rho})^{2} e^{-R^{\sigma '}} $$

Using Chebyshev' s inequality we get: 
$$ \mbox{mes } [ x \in \mathbb{T}  :  | \mbox{(IV)} | > \frac{S}{\rho} R^{-a} ]
< (\frac{S}{\rho} R^{-a})^{- 2} \cdot (\frac{S}{\rho})^{2}  e^{-R^{\sigma '}}
= R^{2 a}  e^{-R^{\sigma '}} $$ 
Therefore, since $ \sigma <  \sigma ' $, we have :
\begin{equation}\label{sum4}
\mbox{mes } [ x \in \mathbb{T}  :  | \mbox{(IV)} | > \frac{S}{\rho} R^{-a} ] 
<  e^{-R^{\sigma }}
\end{equation} 
provided $ R \geq R_0 (a, \sigma) $.

The estimate (\ref{shiftldt}) follows now from (\ref{sum123}) and (\ref{sum4}).
\end{proof}

Notice that in the estimate (\ref{shiftldt}) which we have just proved, 
the greater the power $a$ is, the stronger the estimate becomes. For an arbitrary subharmonic function $u (x)$,\, $a = 1 \, - $ \, is probably optimal.
 
Also notice that as $a \rightarrow 1$  we have $\sigma \rightarrow 0$ 
 and $ p \rightarrow \infty$, so we need higher and higher order averages 
to get (\ref{shiftldt}).
 
On the other hand, we can work with first-order averages $(p = 1)$ as long 
as we only need $ a < \frac{1}{2} $ (say $ a = \frac{1}{3} $) and $\sigma  < 1 - a $ 
(say $\sigma = \frac{1}{3} $). In particular we get : 
\begin{corollary}\label{cshifts}
 Let  $ u : \mathbb{T} \rightarrow \mathbb{R} $, $ \omega \in DC_{\kappa} $ and $ \rho > 0 $. Assume that $ u (x) $ has a subharmonic extension  $ u (z) $ 
to the strip  $ [ | \Im z | < \rho ] $  so that $ | u (z) | \leq  S $  for all 
$ z \in  [ | \Im z | < \rho ] $. Then, for $ R \geq R_0 (\kappa) $ we have: 
\begin{equation}\label{shiftldt2} 
\mbox{mes } [ x \in \mathbb{T}  : | \frac{1}{R} \sum_{j = 0}^{R - 1} u (x + j \omega) - < u > | > \frac{S}{\rho} R^{- 1/3} ] < e^{-R^{1/3}}
\end{equation}
\end{corollary}
 
\section{Large deviation theorem: the case $ s < 2 $ }

\begin{theorem}\label{LDT<2} 
Consider the Schr\"{o}dinger operator (\ref{op2}):
$$ H_{\omega} (x)  :=  - \Delta +   v( x + n \omega ) \delta_{n,n'} $$
with $ v \in G^{s} (\mathbb{T})$  where $ 1 < s < 2 $, and  $ \omega \in DC_{\kappa} $  for some $ \kappa > 0 $. 

Then, for every energy $ E \in \mathbb{R}$,
\begin{equation}\label{ldt<2} 
\mbox{mes } [ x \in \mathbb{T}  : | \frac{1}{N} \log || M_{N} (x, E) || - L_{N} (E) | > N^{- \tau} ] <  e^{-N^{\sigma }}
\end{equation}
for some positive constants $\tau$, $ \sigma $ which depend on $s$, and for $ N \geq  N_0 (\kappa, v, s)$.
\end{theorem}
 
\begin{proof}
We fix the energy $E$ (the estimates will not depend on it), so we can 
drop it from notations. The order $s$ of the Gevrey class satisfies $s \in (1 , 2)$, 
so there is $\delta \in (0 , 1) $, $ s  <  1 + \delta $. Put $ b := \frac{\delta}{s - 1} $ $ > 1 $, and recall Remark 2.1 (note that here the disorder $\lambda $ is fixed):
 
For all $ x \in \mathbb{T} $ and for $N$ large enough, $ N \geq N_0 (v, s) $, if we consider 
$ u_N (x) := \frac{1}{N} \log || \tilde{M}_{N} (x) || $ and  $< u_N > := \int_{\mathbb{T}} u_N (x) dx$, then: 
\begin{equation}\label{subst1} 
\left| \, \frac{1}{N} \log || M_{N} (x) || -  u_N (x) \right| <  e^{-N^{b}}
\end{equation}
\begin{equation}\label{subst2}
 \left| \, L_N  -  < u_N > \right|  <  e^{-N^{b}}
 \end{equation}
 
Moreover, $u_N (x)$ extends to a subharmonic function $u_N (z)$ on the strip $ [ | \Im z | < \rho_{N} ] $, 
$ \rho_{N} \approx N^{- \delta} $, so that $ | u_N (z) | \leq  S $ for all $z$ in the strip, and uniformly in $N$ (here $S$ depends only on $v$, namely on $\sup_{x \in \mathbb{T}} | v (x) | = B$).
 
We will apply Theorem \ref{shifts} for $ u (x) = u_N (x) $ as follows.

Choose $ a \in  (\delta , 1)$, 
$ \sigma \in  (0 , 1 - a) $, $ p \in  \mathbb{N} $, $ p > \frac{a}{1 - a} $; take $R = N^{1 - \epsilon} $, where $ \epsilon  > 0 $ is sufficiently small. Then $ \frac{S}{\rho_N} R^{- a} \approx S N^{\delta} N^{- (1 - \epsilon) a} \approx N^{- c} $, where $c = (1 - \epsilon) a - \delta  >  0 $ if  $\epsilon$ is 
small enough. 

Theorem \ref{shifts} then implies:
\begin{equation}\label{shiftldtN}
\hspace{-.1in} \mbox{mes }[ x\in \mathbb{T} : | \frac{1}{R^p} \sum_{j = 0}^{p ( R - 1)} c_{R}^{p} (j) \cdot u_N (x + j \omega) - < u_N > | >  N^{- c} ] < e^{- N^{\sigma_1}}
\end{equation}
for $N$ (therefore $R$) large enough, $N \geq N_0 (\kappa, s)$. The positive constants $c $ and $\sigma _1 $ (
$=  (1 - \epsilon) \sigma $) depend only on $s$. 

We now have to compare $\frac{1}{R^p} \sum_{j = 0}^{p ( R - 1)} c_{R}^{p} (j) \cdot u_N (x + j \omega) $ 
and $ u_N (x) $. Recalling Remark 2.2, for any $ x \in \mathbb{T} $  and for every $N$, (\ref{mshift}) holds, so:
$$  \left| \frac{1}{N} \log || M_{N} (x) || - \frac{1}{N} \log || M_{N} (x + \omega) || \, \right| < \frac{S}{N} $$
Combining this with (\ref{subst1}), we get that for any $x$, $\omega$ and for $N$ large enough, 
$$ | u_N (x) - u_N (x + \omega) | < \frac{S}{N} $$
Therefore, for every integer $j$ : 
$$ | u_N (x) - u_N (x  +  j \omega) |  < \frac{S |j|}{N} $$
It follows that for any $ x, \omega  \in \mathbb{T} $, and for $N$ large enough 
\begin{equation}\label{avshift}
 | u_N (x) - \frac{1}{R^p} \sum_{j = 0}^{p ( R - 1)} c_{R}^{p} (j) \cdot u_N (x + j \omega) | \leq \frac{S p (R - 1)}{N} \lesssim \frac{R}{N} = N^{- \epsilon}
\end{equation}
The estimate (\ref{ldt<2}) (with $ 0 < \tau < \min \{ c , \epsilon \} $) follows from (\ref{shiftldtN}) and (\ref{avshift}). 
\end{proof} 
\begin{remark}\label{remcts}\rm{
Regarding the constants $\tau$, $\sigma$ in (\ref{ldt<2}), notice from the above proof that $\tau$, $ \sigma$ $ \rightarrow 0 $ as $s \rightarrow 2$.}
\end{remark}

\section{ Large deviation theorem: the general case}
 
We prove the LDT for the Schr\"{o}dinger operator (\ref{op1}) where the function $v \in G^{s} (\mathbb{T})$ with $ s > 1 $ arbitrarily large  and  the frequency $ \omega \in DC_{\kappa} $  for some $ \kappa > 0 $.

We use Remark 2.1. Choose $\delta > 0 $ so that $ s < 1 + \delta $ (say $\delta := 2(s - 1) $), and 
consider at every scale $N$ the corresponding truncation $v_N (x)$ of the function 
$v (x) $. This will give, for every set of parameters $\omega$, $\lambda$, $E$, the transfer matrix 
substitute $ \tilde{M}_N (x)$ and the function $u_N (x) := \frac{1}{N} \log || \tilde{M}_{N} (x) || $ which extends on the strip of width $\approx N^{- \delta}$ to a subharmonic function $ u_N (z)$ satisfying 
$ | u_N (z) | \leq  S(\lambda) $ uniformly in $N$. Moreover, if $ N \geq C S( \lambda) $, from (\ref{aproxu}) and (\ref{aprox<u>}) we get
\begin{equation}\label{Aproxu}  
\left| \, \frac{1}{N} \log || M_{N} (x) || - u_N (x) \, \right| < e^{-c N^{2}}
\end{equation}
\begin{equation}\label{Aprox<u>} 
 \left| \, L_N - < u_N  >  \, \right|  <  e^{-c N^{2}}
\end{equation}

Notice that since $s$ is arbitrarily large, so is $\delta $, and we cannot use the method from Section 4 to prove the LDT. We will follow the ideas in \cite{BGS} and use instead the avalanche principle to boost the estimates on $u_N (x)$ given by  Corollary 3.1. 
For the reader's convenience, we reproduce here the  statement of the 
avalanche principle (see \cite{GS} for the proof) :

\begin{proposition} \label{avalanche} 
Let $A_1, \ldots , A_n$ be a sequence of arbitrary $SL_2 (\mathbb{R})$
matrices. Suppose that 
\begin{equation}\label{aval1}
\min_{1 \leq j \leq n} || A_j || \geq \mu  \geq n
\end{equation}
\begin{equation}\label{aval2} 
 \max_{1 \leq j \leq n} [ \log || A_{j+1} ||  +  \log || A_j  ||  -  \log || A_{j+1}  A_{j} || ]  \leq \frac{1}{2} \log \mu 
\end{equation}  
Then
\begin{equation}\label{aval3} 
 | \,  \log || A_n  \cdot \ldots \cdot A_1  || + \sum_{j = 2}^{n - 1}  \log || A_j ||  - \log || A_{j+1}  A_{j} || \,| \leq C \frac{n}{\mu} 
 \end{equation}
\end{proposition}

We prove the LDT by induction on the scale $N$. The initial condition step 
of the induction follows from the transversality condition only - we don't need any regularity condition here, but just the ``non singularity'' of $v(x)$. 
The regularity condition (\ref{GC}) is needed for the inductive step.

\begin{lemma}{(The inductive step)}\label{ind} 

The data is the following: a function $v \in G^{s} (\mathbb{T})$, $ s > 1 $, $\delta  := 2(s - 1)$, $ D := \delta + 3 $, $ A := \max \{12 \cdot \delta, \, 2 \}  $; some fixed parameters $ \omega $, $\lambda $, $E$ such that $ \omega \in DC_{\kappa} $  for some $ \kappa > 0 $, $ | E | \leq | \lambda | B + 2 $ ( where $ \sup_{ x \in \mathbb{T}} |v (x) | \leq B $ as in Section 2); a fixed number $\gamma > \frac{1}{4}$.
 
Assume $N_0 $, the small scale, is a sufficiently large integer,
$N_0 \geq  N_{0 0} (s, \kappa) $, so that Corollary 3.1 applies at this scale and so that different 
powers and exponentials of $N_0$ behave as they are suppose to do asymptotically, 
e.g. $N_{0}^{A} \ll e^{\frac{9}{40} N_0} $ etc.

Assume (\ref{Aproxu}) holds at scale $N_0$, that is, $N_0$ has to satisfy 
\begin{equation}\label{scale2}
N_0 \geq C S(\lambda) \hspace{.1in} \Leftrightarrow \hspace{.1in} 
| \lambda | \leq e^{c N_0}
\end{equation}
Take $N$, the large scale, so that : 
\begin{equation}\label{scale1} 
N_{0}^{A} \leq N \leq  e^{\frac{9 \gamma}{10} N_0} 
\end{equation}

Suppose the following hold :
 \begin{equation}\label{indhyp1}
\mbox{mes } [ x \in \mathbb{T} : | \frac{1}{N_0} \log || M_{N_0} (x, \lambda, E ) || - L_{N_0} (\lambda, E ) | >  \frac{\gamma}{10} S (\lambda) ]  <  N^{- D}
\end{equation}
\begin{equation}\label{indhyp2}
 \mbox{mes } [ x \in \mathbb{T} : | \frac{1}{2 N_0} \log || M_{2 N_0} (x, \lambda, E ) || - L_{2 N_0} (\lambda, E ) | >  \frac{\gamma}{10}S (\lambda) ]  <  N^{- D}
\end{equation}
\begin{equation}\label{indhyp3}
 L_{ N_0} (\lambda, E ), \, L_{2 N_0} (\lambda, E ) \geq \gamma S(\lambda)
 \end{equation}
\begin{equation}\label{indhyp4}
 L_{ N_0} (\lambda, E ) -  L_{2 N_0} (\lambda, E ) \leq  \frac{\gamma}{40}S (\lambda) 
\end{equation}

Then there are absolute constants $ c, \, C_0 > 0 $ so that
 \begin{equation}\label{indc1}
L_{N} (\lambda, E ) \geq \gamma S(\lambda) - 2 (  L_{ N_0} (\lambda, E ) -  L_{2 N_0} (\lambda, E ) ) - C_{0} S (\lambda) N_{0} N^{-1} 
\end{equation}
\begin{equation}\label{indc2}
L_{ N} (\lambda, E ) -  L_{2 N} (\lambda, E ) \leq  C_{0} S (\lambda) N_{0} N^{-1}
\end{equation}
\begin{equation}\label{indc3}
 \mbox{mes } [ x \in \mathbb{T} : | \frac{1}{N} \log || M_{N} (x, \lambda, E ) || - L_{N} (\lambda, E ) | >  S( \lambda )  N^{-1/10} ]  <   e^{- c {N^{1/10}}}
\end{equation}

\end{lemma}

\begin{proof}
The parameters $\omega$, $ \lambda$, $E$ are fixed, so they can be suppressed from 
the notations. For instance $ M_N (x) = M_N (x, \omega, \lambda, E) $, $ S(\lambda) = S $ etc.
 
We can assume, without loss of generality, that $N$ is a multiple of $N_0$, that is, $ N = n \cdot N_0 $.
 
Indeed, if $ N = n \cdot N_0 + r $, $0 \leq r <  N_0 $, then 
\begin{equation}\label{N/N_0}
 | \frac{1}{N} \log || M_{N} (x) || - \frac{1}{n \cdot N_0} \log || M_{n \cdot N_0} (x) || \, | 
 \leq 2 S N_0 N^{- 1}    
\end{equation}
 
Therefore, if we prove (\ref{indc1}), (\ref{indc2}), (\ref{indc3}) at scale $ n \cdot N_0 $, then they hold 
at scale $N$ too.
 
To prove (\ref{N/N_0}), first notice that $ M_N (x) = B (x) \cdot M_{n \cdot N_0} (x) $, where 
$$ B (x) := \prod_{j=N}^{n \cdot N_0 + 1} A (T^j x) = \prod_{j=n \cdot N_0 + r}^{n \cdot N_0 + 1} A (T^j x)$$
so
$$ || B (x) ||   \leq e^{r \cdot S} \leq e^{N_0 \cdot S} \mbox{  and   } \, || B (x) ^{-1} ||    \leq e^{r \cdot S} \leq e^{N_0 \cdot S} $$

Since $ ||M_{n \cdot N_0} (x) || \geq 1  $  and  $ ||M_{N} (x) || \geq 1  $, it follows that:
$$  \frac{1}{N} \log || M_{N} (x) || - \frac{1}{n \cdot N_0} \log || M_{n \cdot N_0} (x) ||  = 
 \frac{1}{n \cdot N_0} \log \frac{|| M_{N} (x) ||^{\frac{n \cdot N_0}{N}}}{|| M_{n \cdot N_0} (x) ||}  \leq $$
$$ \leq   \frac{1}{n \cdot N_0} \log \frac{|| B (x) ||^{\frac{n \cdot N_0}{N}} \cdot || M_{n \cdot N_0} (x) ||^{\frac{n \cdot N_0}{N}}}{|| M_{n \cdot N_0} (x) ||}  \leq $$
$$ \leq  \frac{1}{n \cdot N_0} \log \, (e^{N_0 S})^{\frac{n \cdot N_0}{N}} = S N_0 N^{- 1} $$

Similarly
$$ \frac{1}{n \cdot N_0} \log || M_{n \cdot N_0} (x) || -  \frac{1}{N} \log || M_{N} (x) || =
\frac{1}{n \cdot N_0} \log \frac{|| M_{n \cdot N_0} (x)||}{ || M_{N} (x) ||^{\frac{n \cdot N_0}{N}}} = $$
$$ = \frac{1}{n \cdot N_0} \, \log \; [ \, \left( \frac{|| M_{n \cdot N_0} (x) ||}{|| M_N (x) ||} \right)^{\frac{n \cdot N_0}{N}} \cdot \, || M_{n \cdot N_0} (x) ||^{\frac{r}{N}} \, ] \leq  $$ 
$$ \leq \frac{1}{n \cdot N_0} \,  \log \; [ \,
|| ( B (x) )^{-1} ||^{\frac{n \cdot N_0}{N}} \cdot || M_{n \cdot N_0} (x) ||^{\frac{r}{N}} \, ] \leq $$
$$ \leq \frac{1}{n \cdot N_0} \, \log \; [ \, (e^{ N_0 S})^{\frac{n \cdot N_0}{N}} \cdot (e^{n  N_0 S})^{\frac{N_0}{N}} ] \, = 2 S N_0 N^{- 1} $$
and the inequality (\ref{N/N_0}) now follows.

Denote the set in (\ref{indhyp1}) by $B_{N_0}$ and similarly the set in (\ref{indhyp2}) by $B_{2 N_0}$.\\
If $ x \notin B_{N_0} $ then using (\ref{indhyp1}), (\ref{indhyp3}) we get 
$$ || M_{N_0} (x) || >  e^{- \frac{\gamma}{10} S  N_{0} +  L_{N_0} \cdot \, N_0 } \geq e^{\frac{9 \gamma}{10} S  N_{0} } =: \mu \geq N $$
so
\begin{equation}\label{avalp1}
 || M_{N_0} (x) || \geq \mu \geq n \hspace{.2in} \mbox{ if } x \notin B_{N_0}
\end{equation} 

For $ 1 \leq j \leq n = \frac{N}{N_0} $ consider $ A_{j} = A_{j} (x) := M_{N_0} (T^{(j-1) N_0} x) $. Then (\ref{avalp1}) implies
\begin{equation}\label{avalpp1}
\min_{1 \leq j \leq n} ||  A_{j} (x) || \geq \mu  \hspace{.2in} \mbox{ for all } 
 x \notin \bigcup _{j=0}^{n} T^{ - j N_0}  B_{N_0}
\end{equation}

Since $ A_{j+1} (x) \cdot A_{j} (x) =  M_{2 N_0} (T^{(j-1) N_0} x) $, using (\ref{indhyp1}), (\ref{indhyp2}), (\ref{indhyp4}), for  \\ $ x \notin  \bigcup _{j=0}^{n} (T^{ - j N_0}  B_{N_0}) \cup   \bigcup _{j=0}^{n} (T^{ - j N_0}  B_{2 N_0}) $ (which is a set of measure \\ $ < 2 N^{- D} \cdot N = 2 N^{- D + 1} $), we have :
$$ \log ||  A_{j+1} (x) || +  \log ||  A_{j} (x) || - \log ||  A_{j+1}(x) \cdot A_{j}(x) ||  = $$ $$ = 
\log || M_{N_0} (T^{j N_0} x) || + \log || M_{N_0} (T^{(j-1) N_0} x)|| - \log || M_{2 N_0} (T^{(j-1) N_0} x)||   \leq $$ $$ \leq  N_{0} (  L_{N_0} + \frac{S \gamma}{10} ) +  N_{0} (  L_{N_0} + \frac{S \gamma}{10} ) + 2 N_{0} (\frac{S \gamma}{10} -  L_{2 N_0}) =  $$ $$ =  2 N_{0} (  L_{N_0} -   L_{2 N_0}) +  \frac{4 S \gamma}{10} N_{0} \leq  \frac{9 S \gamma}{20} N_{0} = \frac{1}{2} \log \mu $$
Therefore,
\begin{equation}\label{avalpp2}
 \log ||  A_{j+1}(x) || +  \log ||  A_{j}(x) || - \log ||  A_{j+1}(x) \cdot A_{j}(x) || \leq  \frac{1}{2} \log \mu 
\end{equation}
for $x$ outside a set of measure $ < 2 N^{- D + 1} $.

\medskip

We can now apply the avalanche principle (Proposition \ref{avalanche}) and get:
\begin{equation}\label{avalpp3}
 | \log || A_{n}(x) \cdot ... \cdot A_{1}(x)|| + \sum_{j=2}^{n-1} \log ||A_{j}(x)||  - \sum_{j=1}^{n-1} \log ||A_{j+1}(x) \cdot A_{j}(x)|| \, | \leq C \frac{n}{\mu} 
 \end{equation}
for $x$ outside a set of measure $ < 2 N^{- D + 1} $.

Hence, since $N = n \cdot N_0$ and $ A_{n}(x) \cdot ... \cdot A_{1}(x) =  M_{N} (x)$, we have:
$$| \log ||  M_{N} (x)|| +  \sum_{j=2}^{n-1} \log || M_{N_0} (T^{(j-1) N_0} x)|| -  $$
$$ - \sum_{j=1}^{n-1} \log ||  M_{2 N_0} (T^{(j-1) N_0} x)|| \, |   \leq C \frac{n}{\mu}$$ 
Therefore
\begin{equation}\label{9}
 | \, \frac{1}{N}  \log ||  M_{N} (x)|| +  \frac{1}{n} \sum_{j=2}^{n-1} \frac{1}{N_0} \log || M_{N_0} (T^{(j-1) N_0} x)|| \,  -
 \end{equation}
 $$ - \, \frac{2}{n} \sum_{j=1}^{n-1} \frac{1}{2 N_0} \log ||  M_{2 N_0} (T^{(j-1) N_0} x)|| \, |   \leq  \frac{C}{\mu}
$$

In (\ref{9}) replace $x$ by each of the elements $ \{ x, T x , ... T^{N_{0}-1} x \} $ and then average (add all the $N_0$ inequalities  obtained and divide by $N_0$). We get:
\begin{equation}\label{90}
|  \frac{1}{N_0} \sum_{j=0}^{N_{0}-1} \frac{1}{N} \log || M_{N} (T^{j} x)||  +  
\frac{1}{N} \sum_{j=0}^{N-1} \frac{1}{N_0} \log || M_{N_0} (T^{j} x)|| \, - 
\end{equation}
$$
- \, \frac{2}{N} \sum_{j=0}^{N-1} \frac{1}{2 N_0} \log ||  M_{2 N_0} (T^{j} x)|| \, |   \leq    \frac{C}{\mu}
$$

Using Remark 2.2 we have:
\begin{equation}\label{900}
 | \frac{1}{N}  \log ||  M_{N} (x)|| -  \frac{1}{N_0} \sum_{j=0}^{N_{0}-1} \frac{1}{N} \log || M_{N} (T^{j} x)|| \, | \leq \frac{C S N_0}{N} 
\end{equation}

From (\ref{90}) and (\ref{900}) we get: 

\begin{equation}\label{10}
| \frac{1}{N}  \log ||  M_{N} (x)|| + 
\frac{1}{N} \sum_{j=0}^{N-1} \frac{1}{N_0} \log || M_{N_0} (T^{j} x)|| \, - 
\end{equation}
$$
- \, \frac{2}{N} \sum_{j=0}^{N-1} \frac{1}{2 N_0} \log ||  M_{2 N_0} (T^{j} x)|| \, |   \leq   \frac{C S N_0}{N} + \frac{C}{\mu} \,  \leq  C S N_0 N^{- 1} $$
for  $ x \notin  B_1 := \bigcup _{j=0}^{N} (T^{ - j }  B_{N_0}) \cup   \bigcup _{j=0}^{n} (T^{ - j }  B_{2 N_0})$ where $\mbox{mes }[ B_1 ] < 2 N^{- D + 1}$.

Integrating the left hand side of (\ref{10}) in $x$, we get:
\begin{equation}\label{11}
 |  L_{N} +  L_{N_0} - 2  L_{2 N_0} | < C S N_0 N^{-1}  + 4 S \cdot 2 N^{- D + 1}  <  C_{0} S N_0 N^{-1}
 \end{equation}
$$ \Rightarrow  L_{N} +  L_{N_0} - 2  L_{2 N_0} > -  C_0  S N_0 N^{-1}$$
$$ \Rightarrow   L_{N} >  L_{N_0} -  2 (  L_{ N_0} -  L_{2 N_0} ) - C_{0} S N_{0} N^{-1} >  \gamma S  -  2 (  L_{ N_0} -  L_{2 N_0} ) - C_{0} S N_{0} N^{-1} $$ 
which proves (\ref{indc1}).

Clearly all of the arguments above work for $N$ replaced by $2 N $, so we get the analogue of (\ref{11}) :
\begin{equation}\label{111} 
|  L_{2 N} +  L_{N_0} - 2  L_{2 N_0} |  <  C_{0} S N_0 N^{-1}
\end{equation}
From (\ref{11}) and (\ref{111}) we obtain
$$ L_{N} -  L_{2 N} \leq  C_{0} S N_0 N^{-1} $$
which is exactly (\ref{indc2}).

To prove (\ref{indc3}) consider $ u_{N_0} (x)  := \frac{1}{N_0} \log || \tilde{M}_{N_0} (x) ||$ which extends to  a subharmonic function  $ u_{N_0} (z) $ on the strip $ [ | \Im z| < \rho _{N_0} \approx N_{0}^{- \delta} ] $ so that for $z$ in this strip, $ | u_{N_0} (z) | \leq S $. The same holds for $  u_{2 N_0} (x)$.

Using (\ref{Aproxu}) which holds at scales $N_0$ and $ 2 N_0$ by (\ref{scale2}), we can 'substitute' \\ in (\ref{10})
$ \frac{1}{N_0} \log || M_{N_0} (T^{j} (x)|| $ by $  u_{N_0} (T^j x) $ and  $ \frac{1}{2 N_0} \log || M_{2 N_0} (T^{j} (x)|| $ by \\ $  u_{2 N_0} (T^j x) $  and get, for $ x \notin B_1$: 
\begin{equation}\label{12}
| \frac{1}{N}  \log ||  M_{N} (x)|| +  \frac{1}{N} \sum_{j=0}^{N-1} u_{N_0}(T^{j} x) -  \frac{2}{N} \sum_{j=0}^{N-1}  u_{2 N_0}(T^{j} x) | <  C S N_0 N^{-1}
\end{equation} 

Applying Corollary 3.1 to $ u_{N_0} $ and $ u_{2 N_0}$ we get :
\begin{equation}\label{13}
\mbox{ mes } [ x \in \mathbb{T} :  | \frac{1}{N} \sum_{j=0}^{N-1} u_{N_0}(T^{j} x) - <u_{N_0}> | >  
S \cdot N_{0}^{\delta} \cdot N^{- 1/3 } ] < C  e^{- N^{1/3}}
\end{equation}
\begin{equation}\label{13'}
\mbox{ mes } [ x \in \mathbb{T} :  | \frac{1}{N} \sum_{j=0}^{N-1} u_{2 N_0} (T^{j} x) - <u_{2 N_0}> | > 
 S \cdot N_{0}^{\delta} \cdot N^{- 1/3 } ] < C  e^{- N^{1/3}}
\end{equation}

Denote the union of the two sets in (\ref{13}), (\ref{13'}) by $B_2$. 

Since $N$ satisfies (\ref{scale1}), $$ S \cdot N_{0}^{\delta} \cdot N^{-1/3} 
< S \cdot (N^{1/A})^{\delta} \cdot N^{-1/3}  < S \cdot   N^{-1/4} $$
so from (\ref{12}), (\ref{13}), (\ref{13'}) we get:
\begin{equation}\label{14}
| \, \frac{1}{N}  \log ||  M_{N} (x)||  \, + \,  < u_{N_0} >  \, - \,  2  < u_{2 N_0} >  \, |  
< 
\end{equation}
$$ < C S N_0 N^{-1}  +  S  \cdot   N^{-1/4}  <  2  S  \cdot N^{-1/4}
$$
for $x  \notin B := B_1  \cup  B_2 $, where  $$\mbox{mes } [ B ] < 2 N^{- D + 1} + 2 e^{-N^{1/3}} <  3 N^{- D + 1} <   N^{- D + 2} .$$
Using (\ref{Aprox<u>}) at scales $N_0$, $2 N_0$ and taking into account (\ref{scale1}), (\ref{14}) becomes:
\begin{equation}\label{15}
| \, \frac{1}{N}  \log ||  M_{N} (x)|| \, + \,  L_{N_0}  \, - \,   2 L_{2 N_0} |  < 2 S \cdot N^{-1/4} \, + \, 2 e^{- c N_{0}^2} < 3 S N^{-1/4}  
\end{equation}
provided $x \notin B$.

Combine (\ref{15}) with (\ref{11}) to get:
\begin{equation}\label{16}
 | \, \frac{1}{N}  \log ||  M_{N} (x)|| \, - \,  L_{N} |  <  C_{0} S N_0 N^{-1} + 3 S \cdot N^{-1/4}  <  S \cdot N^{-1/5} 
 \end{equation}
for all $ x \notin B $, where $\mbox{mes }[ B ] <  N^{- D + 2} $.

Notice that (\ref{16}) is not exactly what we need in order to prove the estimate (\ref{indc3}). We have to prove an estimate like (\ref{16}) for $x$ outside an exponentially small set. We will use Corollary 4.10 in \cite{B} (see also Lemma 2.3 in \cite{BGS}) to boost (\ref{16}) to the desired estimate (\ref{indc3}).
We reproduce here, for convenience, the ``rescaled'' result in \cite{B} (we have to take into account the width $\rho$ of the subharmonic extension of $u(x)$) :

\begin{proposition}\label{boost}
Assume $ u = u (x) : \mathbb{T} \rightarrow \mathbb{R} $ has a subharmonic extension $ u (z) $ to the strip $ [ | \Im z | < \rho ] $, $\rho > 0 $, so that $ | u (z) | \leq S $ for all $z$. If
\begin{equation}\label{weak}
 \mbox{ mes } [ x \in \mathbb{T} :  | \, u (x) - < u > | > \epsilon _0 ] < \epsilon _1
 \end{equation}
then, for an absolute constant $c > 0$, 
\begin{equation}\label{strong}
 \mbox{ mes } [ x \in \mathbb{T} :  | \, u (x) - < u > | > \sqrt{\epsilon _0} ] 
 < e^{- c \left( \sqrt{\epsilon _0} + \sqrt{\frac{\epsilon _1 \cdot S}{\epsilon _0 \cdot \rho}} \; \right)^{- 1}}
 \end{equation}
\end{proposition}

\bigskip

From (\ref{16}), using again (\ref{Aproxu}), (\ref{Aprox<u>}) at scale $N$, we get:
 \begin{equation}\label{160}
 \mbox{ mes } [ x \in \mathbb{T} :   | \, u_{N} (x) - <  u_N  > \, | > S \cdot N^{- 1/5} ] < N^{- D + 2}
 \end{equation}
 
We apply Proposition \ref{boost} to $ u (x) := \frac{1}{S} u_N (x) = \frac{1}{S N} \log || \tilde{M}_N (x) || $.
 
The function  $u (x) $ has a subharmonic extension $ u (z) = \frac{1}{S} u_N (z) $ to the strip $ [ | \Im z | < \rho_N ] $ where $ \rho = \rho_N  \approx N^{- \delta} $, so that $ | u (z) | \leq 1 $ on this strip.
 
The estimate (\ref{160}) implies   
 \begin{equation}\label{1600}
 \mbox{ mes } [ x \in \mathbb{T} :   | \, u (x) - <  u  > \, | >  N^{- 1/5} ] < N^{- D + 2}
 \end{equation}
 
Put  $ \epsilon _0 = N^{- 1/5} $, $ \epsilon _1 = N^{- D + 2}$, $S = 1$, $\rho = N^{-\delta} $, so that
 $$ \sqrt{\epsilon _0} + \sqrt{\frac{\epsilon _1 \cdot S}{\epsilon _0 \cdot \rho}}  = 
 N^{- 1/10} + (N^{- D + 2} N^{1/5} N^{\delta})^{1/2} = N^{- 1/10} + N^{- 4/10} < 2 N^{- 1/10} $$
The estimate (\ref{indc3}) follows then from Proposition \ref{boost}.
\end{proof}
  
\begin{remark}\label{whereitfails1} \rm{
The width $\rho$ of the strip of subharmonicity in the estimate (\ref{strong}) is a great obstacle in extending the LDT to operators given by more general functions $v (x)$. 
Indeed, notice that in order to get a decay in (\ref{strong}) we should have $\epsilon_1 = o (\rho) \Leftrightarrow \frac{1}{\rho} =  \frac{1}{\rho_N} =  o (\frac{1}{\epsilon_1} )  = o (N^C)$, for some big $C$.
On the other hand, when the decay of the Fourier coefficients of $v (x)$ is slower then (\ref{fcoef}), to get the correct error in the approximation of $v (x)$ by $v_N (x)$, deg $v_N$ has to dominate any power of $N$. Hence $\frac{1}{\rho_N} \approx \mbox{deg } v_N \gg N^C $ for any $C > 0$.
}
\end{remark}
\begin{remark}\label{whereitfails2} \rm{
Another obstacle, which we believe prevents this approach (by polynomial approximations) to provide the inductive step in the proof of the LDT when $v (x) $ is in a Sobolev space is the following: the decay of the Fourier coefficients of
 a function  $v (x) $ in a Sobolev space is polynomial. Therefore, we need trigonometric polynomials $v_N (x)$ of degree $\gg e^N$ to obtain the exponentially small error in the approximation. Then the width $\rho_N $ of holomorphicity should satisfy  $\frac{1}{\rho_N} \gg e^N$. If we have the LDT at scale $N_0$ and we want to prove it at scale $N_1$, then we need to use Corollary (\ref{cshifts}) for $ u (x) = u_{N_0} (x) $ and $R = N_1$. Therefore we should have $N_1 \gg \frac{1}{\rho_{N_0}} \gg e^{N_0}$, hence $N_1 \gg e^{N_0}$. The next scale $N_2$ should be then $N_2 \gg e^{N_1}$ and so on. To continue the induction, we should prove that at scale $N_1$ we have:
\begin{equation}\label{impossibletoget}
\mbox{ mes } [ x \in \mathbb{T} : |  \frac{1}{N_1}  \log ||  M_{N_1} (x)|| - 
L_{N_1}   | > N_{1}^{-\epsilon} ] < N_2^{- 10} \ll e^{- N_1}
\end{equation} 
But this is far stronger than an estimate of the form 
$\mbox{mes } [...] < e^{- N_1^{\sigma}}$ for some $ \sigma \in (0, 1)$ - and something which does not even hold in the classical large deviation theory in probabilities, which our LDT mimics here.
}
\end{remark}
We will prove the initial condition step from the induction on scales, via large disorder, and using the transversality condition (\ref{TC}). Let's first consider the transversality condition more throughly.
\begin{lemma} \label{compactarg}
Assume $v$ is a smooth, $2 \pi$-periodic function on $\mathbb{R}$. Then $v$ satisfies the transversality condition (\ref{TC}) if and only if 
 \begin{equation}\label{TC'}
 \exists \, m \geq 1 \hspace{.1in} \exists \, c > 0 \hspace{.1in} \mbox{such that} \hspace{.1in} \forall x \in \mathbb{T} : \hspace{.1in} \max_{1 \leq k \leq m} | \partial^{k} v (x) | \geq c
\end{equation}
The constants $m, c$ in (\ref{TC'}) depend only on $v$.
\end{lemma}  
\begin{proof}
Clearly (\ref{TC'}) $ \Rightarrow $  (\ref{TC}). The converse is an easy compactness argument:
$$ \forall \, x \in [0, 2 \pi ] \hspace{.1in} \exists \, m_x \geq 1 \hspace{.1in} \mbox{such that} \hspace{.1in}  | \partial^{m_x} v \, (x) | >  c_x  > 0 $$
$$ \Rightarrow \hspace{.1in} \exists \, r_x > 0 \hspace{.1in} \mbox{so that if } y \in 
(x - r_x , x + r_x ) \hspace{.1in} \mbox{then} \hspace{.1in}  | \partial^{m_x} v \, (y) | \geq c_x  > 0 $$ 
The family $ \{ \, (x - r_x \, , x + r_x ) \, \} _{x \in [0, 2 \pi ] }$ is a covering of $ [0, 2 \pi ]$. Consider a finite subcover $ (x_1 - r_1, x_1 + r_1 ), \ldots , (x_k - r_k , x_k + r_k ) $.\\
Put $ m := \max \{  m_j  \,: \,  1 \leq j \leq k \} $, $ c := \min \{  c_j  \,: \,  1 \leq j \leq k \} $,
 where $ m_j$, $ c_j$ have obvious meanings, and (\ref{TC'}) follows.
\end{proof}

The following lemma is a {\L}ojasiewicz-type inequality (see \cite{L}). A step in its proof is contained in \cite{E} (see Lemma 3 in \cite{E}).

\begin{lemma}\label{Loj}
Assume that $v$ is a smooth function on $[0, 2 \pi ] $ satisfying the  transversality condition (\ref{TC}).
Then for every $t > 0$
\begin{equation}\label{loj} 
\sup_{E \in \mathbb{R}} \mbox{ mes } [ x \in [0, 2 \pi] : \, | v (x) - E | < t ] < C \cdot t^{b}
\end{equation}
where $C, b > 0 $ depend only on $v$. 
 
\end{lemma} 

\begin{proof}
First we show that (\ref{TC'}) implies 
\begin{equation}\label{loj2}
\mbox{ mes } [ x \in [0, 2 \pi] : \, |  \partial v  \, (x) | < \epsilon ] < C \cdot \epsilon^{1/m}
\end{equation}
for some $ C = C (v) > 0 $ and for every $\epsilon > 0 $.

From Lemma \ref{compactarg}, there are  $ m \geq 1$ and $ c > 0$ such that for all $x \in [0, 2 \pi]$ 
$$ \max_{1 \leq k \leq m} | \partial^{k} v (x) | \geq c  $$
 
Let
$$ A := \max_{1 \leq k \leq m + 1} \, \max_{x \in [ 0, 2 \pi]}  | \partial^{k} v (x) | \, < \infty $$

Partition $[0, 2 \pi ] $ in $ \sim \frac{2 A}{c} $
many intervals of length $ < \frac{c}{2 A}$ each, and 
let $I$ be such an interval. 

Fix $\epsilon > 0 $. We can clearly assume $\epsilon < c $. 

Then either $ | \partial v (x) | \geq \epsilon $ for all $ x \in I$, so we are done with the interval $I$, or there is $ x_0 \in I $ so that   $ | \partial v (x_0) | < \epsilon < c $. In this case, for some 
$ 2 \leq j \leq m $, $ | \partial^{j} v (x_0) | \geq c $. Let's say $ j = m$ (this is the worst case, anyway), so
\begin{equation}\label{a}
 | \partial^{m} v (x_0) | \geq c
\end{equation}
If $ x \in I$, then 
\begin{equation}\label{b} 
|  \partial^{m} v (x) -  \partial^{m} v (x_0) | 
\leq \sup_{y \in I}  | \partial^{m + 1} v (y) | \cdot | x - x_0 | 
\leq A \cdot |I| < \frac{c}{2}
\end{equation}
\begin{equation}\label{c} 
\mbox{(\ref{a}) and (\ref{b})}  \Rightarrow \hspace{.1in}  
| \partial^{m} v (x) | \geq \frac{c}{2} \hspace{.1in} \forall \, x \in I
\end{equation}

Let's now analyze  $ \partial^{m - 1} v $ on $I$.
If for some $ x_1 \in I$ we have  $ | \partial^{m - 1} v (x_1) | <
 \epsilon^{1/m} $, then for every 
 $ x \in I$ with $| x - x_1 | > \frac{4}{c} \cdot \epsilon^{1/m} $, there is $ y \in I$ so that 
$$ |  \partial^{m - 1} v (x) -  \partial^{m - 1} v (x_1) | =
| \partial^{m} v (y) | \cdot | x - x_1 | \geq
\frac{c}{2} \cdot \frac{4}{c} \cdot \epsilon^{1/m} = 2 \, \epsilon^{1/m} $$
Therefore there exists an interval $I_1 \subset I$ , 
$ |I_1| \leq \frac{4}{c} \cdot \epsilon^{1/m} $, so that 
\begin{equation}\label{d}
\mbox{ if } x \in I \backslash I_1, \hspace{.1in} \mbox{ then }
 |  \partial^{m - 1} v (x) | \geq  \epsilon^{1/m}
\end{equation}

Now let's analyze $ \partial^{m - 2} v $ on $I \backslash I_1$, which has at most two connected components, $J_1$, $J_2$.
If for some $ x_2 \in J_1$ we have $ | \partial^{m - 2} v (x_2) | <
 \epsilon^{2/m} $, then for every  $ x \in J_1 \subset  I \backslash I_1$ with  $| x - x_2 | > 
\ 2 \,  \epsilon^{1/m} $, there is $ y \in J_1$ so that 
$$ |  \partial^{m - 2} v (x) -  \partial^{m - 2} v (x_2) | =
| \partial^{m - 1} v (y) | \cdot | x - x_2 | \geq
 \epsilon^{1/m} \cdot 2 \epsilon^{1/m} = 2  \epsilon^{2/m} $$
Therefore we get an interval $I_2 \subset J_1 \subset 
 I \backslash I_1 $, such that 
$ |I_2| \leq 2 \, \epsilon^{1/m} $ and  
\begin{equation}\label{e}
\mbox{ if } x \in J_1 \backslash I_2, \hspace{.1in} \mbox{ then }
 |  \partial^{m - 2} v (x) | \geq  \epsilon^{2/m}
\end{equation}
Doing the same for $J_2$, we get   $I_3 \subset J_2 \subset 
 I \backslash I_1 $  such that 
$ |I_3| \leq 2 \, \epsilon^{1/m} $  and  
\begin{equation}\label{f}
\mbox{ if } x \in I \backslash (I_1 \cup I_2 \cup I_3), \hspace{.05in} \mbox{ then }
 |  \partial^{m - 2} v (x) | \geq  \epsilon^{2/m}
\end{equation}

We continue this for $m - 1$ steps (when we end up with $\partial v$).
We obtain $2^{m} - 1$ intervals, each of length
$ \lesssim  \epsilon^{1/m} $ so that outside these intervals 
$$ |  \partial v \, (x) | \geq  \epsilon^{\frac{m - 1}{m}} \geq \epsilon$$
Therefore,
\begin{equation}\label{g}
\mbox{ mes } [ x \in I : \, |  \partial v  \, (x) | < \epsilon ] < C_1  \cdot \epsilon^{1/m}
\end{equation}
where $C_1$ is a constant which depends on $m$ and $c$.
But then
$$ \mbox{ mes } [ x \in [0, 2 \pi] : \, |  \partial v  \, (x) | < \epsilon ] < \frac{2 A}{c} \cdot C_1  \cdot \epsilon^{1/m} <  C \cdot \epsilon^{1/m}$$
so (\ref{loj2}) is proved.
We are now ready to prove (\ref{loj}).

For $E \in \mathbb{R}$ arbitrarily fixed, we have:
$$  [ x  : \, | v (x) - E | < t \, ]  \subset [ x:  |  \partial v  \, (x) | < \, t^{1/4} \, ] \, \cup $$ $$ \cup \,  [ x : \, |  \partial v  \, (x) | \geq \,  t^{1/4} \, \mbox{and } \, | v (x) - E | < t\,  ] 
 =: E_1 \cup E_2 $$

Using (\ref{loj2}) with $ \epsilon = t^{1/4} $ we get 
$$\mbox{ mes } [ E_1 ] < C \, t^{\frac{1}{4 m}} $$
Now we estimate the measure of $E_2$.
Let $$ A :=  \max_{x \in [ 0, 2 \pi]}  | \partial^{2} v (x) | $$
Partition $[0, 2 \pi ] $ in $ \sim t^{- 1/4} $
many intervals of length $ < \frac{1}{2 A}  t^{1/4} $ each and 
let $I$ be such an interval. If $I \cap E_2 = \emptyset$, then we are done with the interval $I$. Otherwise, let $ x_1 \in I \cap E_2 $, so 
$ | \partial v \, (x_1) | \geq   t^{1/4} $.
If $ x \in I$, then 
$$ | \partial v (x) - \partial v (x_1) | 
\leq \sup_{y \in I} | \partial^{2} v (y) | \cdot | x - x_1 | \leq A \cdot |I| < \frac{1}{2}  t^{1/4} $$

Therefore, if  $ x \in I$, then $  | \partial v (x) | >  \frac{1}{2}  t^{1/4} $.

Now let $ x \in I$ with $ | x - x_1 | > 4  t^{3/4}$. For some $ y \in I$ we have:
$$ | ( v (x) - E ) -  ( v (x_1) - E ) | = | v(x) - v (x_1) | =  
| \partial v (y) | \cdot |x - x_1| 
\geq \frac{1}{2} t^{1/4} \cdot  4  t^{3/4} = 2 t $$
But $ x_1 \in E_2$, so $ |   v (x_1) - E  | < t $.

Therefore,  if  $ x \in I$ with $ | x - x_1 | > 4  t^{3/4}$, then  
 $ |   v (x) - E  | > t $, so $ x \notin E_2$.
 
It follows that $ \mbox{ mes } [ I \cap E_2 ] \leq 4  t^{3/4} $, so
 $ \mbox{ mes } [ E_2 ] \lesssim  t^{- 1/4} \cdot  t^{3/4} =  t^{1/2}$
and the inequality (\ref{loj}) is proved.
\end{proof}

\begin{lemma}{(The initial condition step)} \label{Step1}

Assume  that $ v $ is smooth and that $v$ satisfies the transversality condition (\ref{TC}). Then there are positive constants $ \lambda_1$, $B$ which depend on $v$ and $s$ so that
for all $N$ and for all $ \lambda$ subject to: 
$ | \lambda |  \geq \max \{ \lambda _1, N^B \} $ we have:
\begin{equation}\label{step1}
\sup_{E} \mbox{ mes } [ x \in \mathbb{T} : 
\, | \,  \frac{1}{N} \log || M_{N} (x, \lambda, E) || - L_N (\lambda, E) \, | 
> \frac{1}{20} \, S(\lambda) ]  < N^{- A^2 \cdot D}
\end{equation}
where $A$ and $D$ are the constants defined in Lemma \ref{ind} (they depend on $s$).

Furthermore, for these $\lambda$, $N$ and for all $E$ we have:
\begin{equation}\label{step10} 
 L_N (\lambda, E) \geq \frac{1}{2} \, S(\lambda)
\end{equation}
\begin{equation}\label{step100}
 L_N (\lambda, E) -  L_{2 N} (\lambda, E) \leq  \frac{1}{80} \, S(\lambda)
\end{equation}
\end{lemma}

\begin{proof}
This statement is the analogue of Lemma 2.10 in \cite{BGS}. The model considered in \cite{BGS} is the skew-shift, and the potential $v$ is assumed real analytic. The only fact about the analyticity of $v$ used in the proof of Lemma 2.10 in \cite{BGS} is the {\L}ojasiewicz inequality (\ref{Loj}). We have proved this inequality assuming (only) the transversality condition on $v$ (see Lemma \ref{Loj}). Therefore, the proof of our result, Lemma \ref{Step1} is completely analogous to the proof of Lemma 2.10 in the aforementioned paper. We skip this argument here, but refer the reader to \cite{BGS}.

It should be noted that in \cite{BGS} the measure of the set in (\ref{step1}) is shown to be $ < N^{- 50}$. Here we want it to be $ < N^{- A^2 \cdot D}$. Of course, this will not make any difference in the proof, we just have to choose the power $B$ even larger, depending on $A$ and  $D$.
\end{proof}

We are now ready to prove the main result of this section - the LDT and the positivity of the Lyapunov exponent (statement (P) in Theorem \ref{main}).

\begin{theorem}\label{LDT}

Consider the Schr\"{o}dinger operator (\ref{op1}):
$$ H_{\omega, \lambda} (x)  :=  - \Delta + \lambda \,  v( x + n \omega ) \delta_{n,n'} $$
where the potential $ v \in G^{s} (\mathbb{T})$, $ s > 1 $, and $v$ satisfies the transversality condition (\ref{TC}). Assume that the frequency  $ \omega \in DC_{\kappa} $  for some $ \kappa > 0 $.

There exists $ \lambda_{0} = \lambda_{0} ( v, \kappa ) $ so that for every fixed $\lambda $ with $ | \lambda | \geq \lambda_{0} $ and for every energy $E$, we have:
\begin{equation}\label{ldt}
 \mbox{mes } [ x \in \mathbb{T} : | \frac{1}{N} \log || M_{N} (x, \lambda, E ) || - L_{N} (\lambda, E ) | >  N^{-\tau} ]  <   e^{- {N^{\sigma}}}
\end{equation}
for some absolute constants $ \tau, \sigma  > 0 $, and for all
$ N \geq  N_{0} (\lambda, \kappa, v, s) $.
   
Furthermore, for every such $\omega$, $\lambda$ and for all energies $E \in \mathbb{R}$ we have:
\begin{equation}\label{poslyap}
 L_{\omega, \lambda} (E) \geq \frac{1}{4} \log | \lambda | > 0
\end{equation}
\end{theorem}
\begin{proof}
We show that there exists $ \lambda_0 = \lambda_0 (\kappa, v, s)$ such that if $\lambda$ is fixed, $ | \lambda | \geq  \lambda_0 $ and if 
$ N \geq N_0 (\lambda, \kappa, v, s)$ then for some absolute constant $ c > 0$ we have :
\begin{equation}\label{ldt'}
\mbox{mes } [ x \in \mathbb{T} : | \frac{1}{N} \log || M_{N} (x) || - 
L_{N} | >  S  N^{-1/10} ]  <   e^{- c {N^{1/10}}}
\end{equation}

Clearly (\ref{ldt'}) $ \Rightarrow $ (\ref{ldt}).

Take $N_0$ large enough,
$N_0 \geq  N_{0 0} (s, \kappa) $, so that Corollary 3.1 applies at this scale and so that different 
powers and exponentials of $N_0$ behave as they are suppose to do asymptotically, 
e.g  $N_{0}^{A^2} \ll e^{\frac{9}{40} N_0} $, $ (2 N_0)^B \ll e^{c N_0}$, 
$ N_0^{A^2 \cdot D} \ll  e^{ c {N_0^{1/10}}} $ etc., where $ A, B, c, D $ are the constants introduced earlier in this section (they depend only on $s$).

$N_0$ will be the scale at which we use the initial step - Lemma \ref{ind}.
This lemma gives $ \lambda_1 $, $B$ $ > 0$ such that for every $\lambda$ with 
$ | \lambda |  \geq \max \{ \lambda _1, N_0{^B} \} $ and for $N$ with 
$N_0^{A} \leq N  \leq N_0^{A^2} $, (\ref{indhyp1}), (\ref{indhyp3}), 
(\ref{indhyp4}) hold at scale $N_0$. We also need (\ref{indhyp2}), so just choose $\lambda$ with $ | \lambda |  \geq \max \{ \lambda _1, (2 N_0)^B  \} $.

In order to apply Lemma \ref{ind} to move to a larger scale, we also have to satisfy the condition (\ref{scale2}): $ | \lambda | \leq e^{c N_0} $.

Therefore, in order to use both the initial step and the inductive step at small scale $N_0$ and with disorder $\lambda$, the numbers $N_0$, $\lambda$ have to satisfy:
\begin{equation}\label{alfa}    
(2 N_0)^B \leq  | \lambda |  \leq e^{c N_0} 
\end{equation}
\begin{equation}\label{beta}
N_0 \geq N_{0 0}
\end{equation}
\begin{equation}\label{gama} 
 | \lambda |  \geq  \lambda_1
\end{equation}

But we want to prove the LDT for every disorder $  \lambda $ large enough, 
$  | \lambda |  \geq \lambda_0 $ (not just for $\lambda$ in a bounded interval). 

The trick is to choose first $   \lambda $ large enough, and then to pick 
$ N_0 = N_0 ( \lambda)  \geq   N_{0 0} $ such that (\ref{alfa}) holds. This is possible because:
there is $ \lambda_2 > 0 $ such that for all $ \lambda $ with  $ | \lambda | \geq \lambda_2 $, there exists $ N_0 = N_0 (\lambda) \geq N_{0 0} $ so that (\ref{alfa}) holds . 

Indeed, (\ref{alfa}) $ \Leftrightarrow $ 
$ \frac{1}{c} \log  | \lambda | \leq N_0 \leq  \frac{1}{2} | \lambda |^{1/B}$

We can find $ \lambda_2 $ large enough, $ \lambda_2 =  \lambda_2 (\kappa, v, s)$, so that if $ | \lambda | \geq   \lambda_2 $, then 
$$  \frac{1}{c} \log  | \lambda | \geq   N_{0 0} \, \mbox{  and  } \,  
 \frac{1}{c} \log  | \lambda | \ll   \frac{1}{2} | \lambda |^{1/B}$$
 
It follows that for every such $\lambda$ we can pick $  N_0 = N_0 ( \lambda)$ so that $ \frac{1}{c} \log  | \lambda | \leq N_0 \leq  \frac{1}{2} | \lambda |^{1/B}$, thus (\ref{alfa}), (\ref{beta}), (\ref{gama}) hold.

Now we can start the proof of (\ref{ldt'}). Let
 $ \lambda_0 := \max \{  \lambda_1 ,  \lambda_2 \} $. Fix $  \lambda $ with 
$ | \lambda | \geq  \lambda_0 $.

To start off the induction, choose the initial scale $N_0$ so that  
 $ \frac{1}{c} \log  | \lambda | \leq N_0 \leq  \frac{1}{2} | \lambda |^{1/B}$.
Therefore (\ref{alfa}), (\ref{beta}), (\ref{gama}) hold, and we can apply Lemma \ref{Step1} and Lemma \ref{ind} with small scale $N_0$ and disorder $\lambda$.

If the large scale $N$ is such that
$$ N_0^{A} \leq N \leq  N_0^{A^2} ( \leq  e^{\frac{9}{40} N_0}) $$
from Lemma \ref{Step1} we get
$$ \mbox{mes } [ x \in \mathbb{T} : | \frac{1}{N_0} \log || M_{N_0} (x) || - 
L_{N_0}  | >  \frac{1}{20} S  ]  <  N_0^{- A^2 \cdot D} \leq  N^{- D} $$
$$ \mbox{mes } [ x \in \mathbb{T} : | \frac{1}{2 N_0} \log || M_{2 N_0} (x) || -L_{2 N_0}  | >  \frac{1}{20} S  ]  <   N^{- D} $$
$$ L_{ N_0}, \, L_{2 N_0} \geq \frac{1}{2} S $$
$$  L_{ N_0} -  L_{2 N_0} \leq  \frac{1}{80} \, S $$

Thus, (\ref{indhyp1}) - (\ref{indhyp4}) hold at scale $N_0$ with 
$ \gamma = \gamma_0 = \frac{1}{2}$. Lemma \ref{ind} implies that for the large scale $N$ in the range $ [ N_0^{A}, N_0^{A^2} ] $ we have:
\begin{equation}\label{fin5}
L_{N}  \geq \gamma_0 S - 2 (  L_{ N_0}  -  L_{2 N_0} ) - C_{0} S N_{0} N^{-1} 
\end{equation}
\begin{equation}\label{fin6}
L_{ N}  -  L_{2 N}  \leq  C_{0} S  N_{0} N^{-1}
\end{equation}
\begin{equation}\label{fin7}
 \mbox{mes } [ x \in \mathbb{T} : | \frac{1}{N} \log || M_{N} (x) || - 
L_{N} | >  S \, N^{-1/10} ]  <   e^{- c {N^{1/10}}}
\end{equation} 
where  $ c, \, C_0 > 0 $  are absolute constants.

From (\ref{fin7}) we get (\ref{ldt'}) for the scale $N$ at least in the range  
 $ [ N_0^{A}, N_0^{A^2} ] $.

If $N_1$ is anywhere in this range, say $N_1 =  N_0^{A}$, then (\ref{fin6}) implies:
$$ L_{ N_1}  -  L_{2 N_1}  \leq  C_{0} S  N_{0} N^{-1} \leq 
C_{0}  N_{0}^{- A + 1} \, S \, ( < \frac{1}{4} \cdot  \frac{1}{40} \cdot S) $$
and combining this with (\ref{fin5}),
$$  L_{N_1}  \geq \gamma_0 \, S - 3  C_{0}  N_{0}^{- A + 1} S =
(\gamma_0 -  3  C_{0}  N_{0}^{- A + 1} ) \cdot S =: \gamma_1 \cdot S $$
where $\gamma_1 := \gamma_0 -  3  C_{0}  N_{0}^{- A + 1} > \frac{1}{2} - 
3 \cdot \frac{1}{4} \cdot  \frac{1}{40} >  \frac{1}{4} $.

Therefore we get
\begin{equation}\label{fin8}
L_{N_1}  \geq \gamma_1 S
\end{equation}
\begin{equation}\label{fin9}
L_{ N_1}  -  L_{2 N_1}  \leq \frac{1}{4} \cdot \frac{1}{40} \cdot S 
\leq \frac{\gamma_1}{40} \cdot S
\end{equation}
Also $2 N_1 = 2 N_0^{A} $ is in the range  $ [ N_0^{A}, N_0^{A^2} ] $ so we get (\ref{fin8}) at scale $2 N_1$:
\begin{equation}\label{fin10}
L_{2 N_1}  \geq \gamma_1 S
\end{equation}
If $N_2$ is the next large scale, so that $ N_1^{A} \leq N_2  \leq  N_1^{A^2} $, then since $  e^{- c {N_{1}^{1/10}}} < N_{1}^{- A^2 \cdot D} \leq N_{2}^{- D}$, (\ref{fin7}) implies 
$$ \mbox{mes } [ x \in \mathbb{T} : | \frac{1}{N_1} \log || M_{N_1} (x) || - 
L_{N_1}  | >  \frac{1}{20} S  ]  <    N_{2}^{- D} $$
$$ \mbox{mes } [ x \in \mathbb{T} : | \frac{1}{2 N_1} \log || M_{2 N_1} (x) || -L_{2 N_1}  | >  \frac{1}{20} S  ]  <   N_{2}^{- D} $$
We can apply Lemma \ref{ind} again, with $N_1$ as the small scale, and $N_2$ as the large one, where $N_2 \in  [ N_1^{A}, N_1^{A^2} ] $. 
From (\ref{indc3}) we get (\ref{ldt'}) at least in the range 
$ [ N_1^{A}, N_1^{A^2} ] \, = \,  [ N_0^{A^2}, N_0^{A^3} ] $, while from   
 (\ref{indc2}), (\ref{indc1}) we get:
$$  L_{ N_2}  -  L_{2 N_2}  \leq  C_{0} S  N_{1} N_{2}^{-1} \leq 
C_{0}  N_{1}^{- A + 1} \, S \, ( < \frac{1}{4} \cdot  \frac{1}{40} \cdot S) $$
$$ L_{N_2}  \geq \gamma_1 S - 2 (  L_{ N_1}  -  L_{2 N_1} ) -
 C_{0} S N_{1} N_{2}^{-1} \geq (\gamma_1 - 3 C_0  N_{1}^{- A + 1}) \cdot S
=: \gamma_2 \cdot S $$
where  $\gamma_2 := \gamma_1 -  3  C_{0}  N_{1}^{- A + 1} =
 \frac{1}{2} -  3  C_{0}  N_{0}^{- A + 1} -  3  C_{0}  N_{0}^{A \cdot (- A + 1)}
> \frac{1}{4} $.

Hence
$ L_{N_2} \geq \gamma_2 \cdot S $ and 
$  L_{N_2} -  L_{2 N_2} \leq \frac{\gamma_2}{40} \cdot S $.

Continuing this inductively, we get (\ref{ldt'}) at every scale $ N \geq N_{0}^A $.
Also, at step $k$, if $ N \in  [ N_{k}^{A}, N_{k}^{A^2} ] $, then
$L_N \geq \gamma_k \cdot S > \frac{1}{4} \cdot S  $ so
$$ L = \inf_{N} L_N \geq \frac{1}{4} \cdot S $$ 
and (\ref{poslyap}) is proved.
\end{proof}

\section{ Proof of the main results. Final remarks}

We are now ready to prove Theorem \ref{main} and Theorem \ref{sec}. We will show - both in the perturbative and the nonperturbative case - that the Lyapunov exponent and the IDS are continuous functions with a certain modulus of continuity (which is the statement (C) in Theorem \ref{main} and Theorem \ref{sec}) and that the operators (\ref{op1}) and (\ref{op2}) satisfy Anderson localization (which is the statement (AL) in  Theorem \ref{main} and Theorem \ref{sec}).

\medskip
 
In Theorem \ref{main} we consider the Schr\"{o}dinger operator (\ref{op1}) : 
$$ H_{\omega, \lambda} (x)  :=  - \Delta +  \lambda v( x + n \omega ) \delta_{n,n'}$$
where the function  $v \in G^{s} (\mathbb{T})$, $s > 1$ and it satisfies the transversality condition (\ref{TC}), 
and the frequency $ \omega \in DC_{\kappa} $,  for some $ \kappa > 0 $.

Based on Theorem \ref{LDT}, there is $\lambda_0 = \lambda_0 (v, \kappa) > 0 $ so that if we fix $\lambda $ with $ | \lambda | > \lambda_0 $ (and we suppress it from notations), 
the following is true for $ H_{\omega} (x) = H_{\omega, \lambda} (x) $:

(i.1) The Lyapunov exponent satisfies 
 $$ L_{\omega} (E) > \frac{1}{4} \log | \lambda | =: c_0 > 0 $$
 for all $ \omega \in DC_{\kappa} $ and  for all energies $E \in \mathbb{R}$.
 
(ii.1) For every energy $E$ we have the estimate (\ref{ldt}):
$$ \mbox{mes } [ x \in \mathbb{T} : | \frac{1}{N} \log || M_{N} (x, E ) || - L_{N} (E ) | >  N^{- \tau} ]  <  e^{- N^{\sigma}} $$
for some universal positive constants $\tau$, $\sigma$ and for $N \geq N_{0, 0} (\kappa, v)$.

\bigskip
 
In Theorem \ref{sec} we consider the Schr\"{o}dinger operator (\ref{op2}) : 
$$ H_{\omega} (x)  :=  - \Delta +  v( x + n \omega ) \delta_{n,n'} $$
where the function $v \in G^{s} (\mathbb{T})$, $s \in (1 , 2) $, and the frequency $ \omega \in DC_{\kappa} $  for some $ \kappa > 0 $.

(i.2) We assume that the Lyapunov exponent of $H_{\omega} (x) $ satisfies 
$$ L_{\omega} (E) >  c_0 > 0 $$ 
for a.e. $ \omega \in DC_{\kappa} $ and for all energies $E$ in some compact interval $I$ (when we prove the statement (C)) or for all energies $E \in \mathbb{R}$ (when we prove the statement (AL)). 

(ii.2) From Theorem \ref{LDT<2} we get the estimate (\ref{ldt<2}): 
$$ \mbox{mes } [ x \in \mathbb{T} : | \frac{1}{N} \log || M_{N} (x, E ) || - L_{N} (E ) | >  N^{- \tau} ]  <  e^{- N^{\sigma}} $$
for all energies $E$, for some positive constants $\tau$, $\sigma$ which depend only on $s$, 
and for all $N \geq N_{0, 0} (\kappa, v) $ (remember that $\sigma$, $\tau$ $ \rightarrow 0$ as $ s \rightarrow 2$).
 
Therefore, we either prove (for Theorem \ref{main}) or we assume 
(for Theorem \ref{sec}) that the Lyapunov exponent is bounded away from $0$. Also, in both cases we have a LDT for the transfer matrices associated 
to (\ref{op1}) and (\ref{op2}) respectively. 

\begin{proof}[Proof of the statement (C)]

Let $N_0$ be any sufficiently large integer, $N_0 \geq N_{0,0} (\kappa, v, s)$ (so that the LDT holds at this scale, $L_{N_0} (E) - L (E)  <  \frac{c_0}{40} $, $N_0^{- \tau} < \frac{c_0}{10} $ etc.).

Then from the LDT we have:
\begin{equation}\label{x1}
\mbox{mes } [ x \in \mathbb{T} : | \frac{1}{N_0} \log || M_{N_0} (x, E ) || - L_{N_0} (E ) | >  N_{0}^{-\tau} ]  <   e^{- {N_{0}^{\sigma}}}
\end{equation}
so for $x$ outside a set of measure $ <  e^{- {N_{0}^{\sigma}}}$,
$$ \frac{1}{N_0} \log || M_{N_0} (x, E ) || > L_{N_0} (E ) - N_{0}^{-\tau} > \frac{9 c_0}{10} $$ 

Choose $0 < \eta < \sigma \, ( < 1) $ and consider the large scale $N \approx e^{(3 S N_0)^{\eta}}$, so \\ $ N_0 \approx \frac{1}{3 S} (\log N)^{1/ \eta} $.

From (\ref{x1}) we get:
\begin{equation}\label{x2}
\mbox{mes } [ x \in \mathbb{T} : | \frac{1}{N_0} \log || M_{N_0} (x, E ) || - L_{N_0} (E ) | >  \frac{c_0}{10} ]  <   e^{- {N_{0}^{\sigma}}} < N^{-3}
\end{equation}

We may clearly assume that (\ref{x2}) holds for $N_0$ replaced by $2 N_0$. Also notice that 
$$ L_{N_0} (E) - L_{2 N_0} (E) < L_{N_0} (E) - L (E) < \frac{c_0}{40}  $$

Therefore, arguing exactly as in the first part of the proof of Lemma \ref{ind} - using the avalanche principle at small scale $N_0$ and large scale $N$ -  we get, after integrating in $x$, (see also (\ref{11}), (\ref{111})):
\begin{equation}\label{y1} 
|  L_{N} (E) +  L_{N_0} (E) - 2  L_{2 N_0} (E) |  <  C S N_0 N^{- 1}
\end{equation}
\begin{equation}\label{y2} 
|  L_{2 N} (E) +  L_{N_0} (E) - 2  L_{2 N_0} (E) |  <  C S N_0 N^{- 1} 
\end{equation}
Hence
\begin{equation}\label{y3} 
|  L_{N} (E) -   L_{2 N} (E) | <   C S N_0 N^{- 1} < C (\log N)^{1/ \eta} N^{- 1}
\end{equation}

Notice that (\ref{y1}), (\ref{y2}), (\ref{y3}) hold for every $N_0$ large enough and for every $N$ such that $N \approx e^{(3 S N_0)^{\eta}}$ (or $ N_0 \approx \frac{1}{3 S} (\log N)^{1/ \eta} $). Therefore, summing over dyadic $N$'s in (\ref{y3}) we get:
\begin{equation}\label{y4} 
|  L_{N} (E) -   L(E) | < C (\log N)^{1/ \eta} N^{- 1}  \approx C N_0 N^{- 1}
\end{equation}
Substituting (\ref{y4}) in (\ref{y1}) we get for every energy $E$:
\begin{equation}\label{z1} 
| L (E) +  L_{N_0} (E) -  2 L_{2 N_0} (E) | <  C N_0 N^{- 1}
\end{equation}

Using  Trotter's formula, we have:
$$ M_{N_0} (x, E) -  M_{N_0} (x, E')  = $$
$$ \sum_{j=1}^{N_0} A (T^{N_0} x, E) \ldots  \,  [A (T^j x, E) - A (T^j x, E')] \,  \ldots A (T x, E')$$
But
$$ A (T^j x, E) - A (T^j x, E') = \left( \begin{array}{cc}
E' - E  &   0  \\
0 &  0 \\  \end{array} \right),$$ 
$$  ||A (T^j x, E)||  \leq  e^{S} \quad \mbox{ for all } E \in I $$
so
$$ || M_{N_0} (x, E) -  M_{N_0} (x, E') || \leq e^{S N_0} \, | E - E'| $$

Therefore, since $ || M_{N_0} (x, E) || \geq 1 $ and  $ || M_{N_0} (x, E') || \geq 1 $, we have 
$$ \left| \, \log || M_{N_0} (x, E) || -  \log || M_{N_0} (x, E) || \, \right| \leq $$ 
$$ \leq
|| M_{N_0} (x, E) -  M_{N_0} (x, E') || \leq e^{S N_0} \, | E - E'| $$

Integrating in $x$ we obtain:
\begin{equation}\label{z2}
| L_{N_0} (E) - L_{N_0} (E') | \leq e^{S N_0} \, | E - E'| 
\end{equation}
Similarly for $2 N_0$ we get:
\begin{equation}\label{z3}
| L_{2 N_0} (E) - L_{2 N_0} (E') | \leq  e^{2 S N_0} \, | E - E'|
\end{equation}
Then from (\ref{z1}), (\ref{z2}), (\ref{z3}) we conclude:
\begin{equation}\label{z4}
| L (E) - L(E') | \leq C N_0 N^{- 1} + 2 e^{2 S N_0} \, | E - E'|
\end{equation}

Set $ | E - E'| =  e^{- 3 S N_0}$, so $ 2 e^{2 S N_0} \, | E - E'| = 2 e^{- S N_0} < N_0 N^{- 1} $

We proved that for any $E$, $E'$ so that $ | E - E'| =  e^{- 3 S N_0}$ (where $N_0$ is any sufficiently large integer), 
$$ | L (E) - L(E') | \leq C N_0 N^{- 1} < C e^{- c (\log \frac{1}{| E - E' |})^{\eta}} $$
where $ C = C(I, \kappa, v, s)$, and $c$ is an absolute positive constant.

This concludes the proof of the continuity of the Lyapunov exponent. The corresponding result for the IDS is obtained by standard methods, using Thouless formula and some elementary properties of the Hilbert transform (see Section 10 in \cite{GS} for more details).
\end{proof} 

\begin{remark} \label{modi} \rm{ The modulus of continuity of the Lyapunov exponent and of the IDS in Theorem \ref{LDT} can be improved to 
\begin{equation} \label{modconi} 
h_{\epsilon} (t) = C_{\epsilon} \, e^{- c | \log t |^{1 - \epsilon}}
\end{equation}
for every $\epsilon > 0$, where $C_{\epsilon} = C (\epsilon, \lambda, \kappa, v, s)$, and $c > 0 $ is an absolute constant.

We sketch the proof of this fact.

First notice that in order to get this stronger modulus of continuity, we only need to prove a sharper version of our LDT, namely the following:
\begin{equation}\label{ldti} 
\mbox{mes } [ x \in \mathbb{T}  : | \frac{1}{N} \log || M_{N} (x, E) || - L_{N} (E) | > N^{- \tau} ] <  e^{-N^{\sigma }}
\end{equation}
for any $ 0 < \sigma < 1$, for some $ \tau > 0$ and for all $N \geq N_0 (\sigma, \tau, \lambda, \kappa, v, s)$. 

This sharper LDT cannot be proved using our result, Theorem \ref{shifts}, on averages of shifts of subharmonic functions. This is because in Theorem \ref{shifts}, in order to obtain sharper estimates, one has to consider higher order averages, which does not fit into the avalanche principle argument used in the proof of the LDT.

However, using Theorem 3.8 in \cite{GS} - where only first order averages are needed to prove sharp estimates on averages of shifts of subharmonic functions - one gets the following:

If $ u = u (x) $ is a function on $ \mathbb{T}$, with a bounded (by $S$) subharmonic extension to a strip of width $\rho$, and if  $ \omega \in DC_{\kappa} $, then
\begin{equation}\label{newshiftldt} 
\mbox{mes }[ x\in \mathbb{T} : | \frac{1}{R} \sum_{j = 0}^{ R - 1} u (x + j \omega) - < u > | > \frac{S}{\rho} R^{- a} ] < 
e^{- R^{\sigma}}
\end{equation}
for any $ 0 < a < 1 $, $ 0 < \sigma  <  1 - a $ and $ R \geq R_0 = R_0 (\kappa, a, \sigma) $.

Using (\ref{newshiftldt}) in our proof of the LDT, instead of Corollary \ref{cshifts}, as well as a modified (optimized) version of Proposition \ref{boost}, one gets (\ref{ldti}), and from there the stronger modulus of continuity (\ref{modconi}).

It should be noted that the modulus of continuity of the Lyapunov exponent and of the IDS given by (\ref{modconi}) is optimal for this method of proof. We don't get H\"{o}lder continuity, although one could expect this to hold, as in the analytic case (see \cite{GS}).
}
\end{remark}

\begin{proof}[Proof of the statement (AL)]
From the LDT (ii.1), (ii.2) - exactly as in \cite{B}, \cite{BG} -
by using Cramer's rule, we obtain the following 'good bounds' 
on the Green's functions $G_{\Lambda} (E, x)$ associated to the operators (\ref{op1}), (\ref{op2}). 

For every $N$ large enough and for every energy $E$, there is a set 
$\Omega_{N} (E) \subset \mathbb{T} $ with  $ \mbox{mes } [ \Omega_{N} (E) ] < e^{- N^{\sigma}} $
so that for any $ x \notin \Omega_{N} (E) $, one of the 
intervals 
$$ \Lambda = \Lambda (x) = [1, N ], [1, N - 1], [2, N ], [2, N - 1] $$ 
will satisfy : 
\begin{equation}\label{green} 
 | G_{\Lambda} (E, x) (n_1 , n_2 ) | < e^{- L(E) | n_1  - n_2 | + N^{1 -}}
 \end{equation} 

Since $ v(x) = \sum_{k \in \mathbb{Z}} \hat{v} (k) e^{i k x} $ and $| \hat{v} (k) | \leq  M e^{- \rho |k|^{1/s}} $, substituting in (\ref{green}) $v (x) $ by $ v_1 (x) := \sum_{|k| \leq C N^{s}} \hat{v} (k) e^{i k x} $,
we can assume that the 'bad set' $\Omega_{N} (E) $
has not only exponentially small measure, but it also has bounded algebraic 
complexity (i.e. it is semi­algebraic of degree  $\, \leq N^{d(s)}$ ). 

The rest of the proof of the localization for the operators (\ref{op1}), (\ref{op2}) uses 
semi­algebraic set theory, and follows exactly the same pattern as the proof 
of the corresponding result for the analytic case. The reader is referred to 
\cite{B}, \cite{BG} (see for instance chapter X in \cite{B}).
\end{proof}
 
\begin{remark} \label{final}\rm{There are some natural problems left unsolved.

One is to extend E. Sorets and T. Spencer result on the positivity of the Lyapunov exponent (see \cite{S-S}) from analytic functions to functions in a Gevrey class $  G^{s} (\mathbb{T})$ - at least for $ s \in (1, 2)$. One also has to decide whether a transversality condition is needed here.

A second problem is to prove nonperturbative localization for the operator (\ref{op1}) when $v \in G^{s} (\mathbb{T})$, where $s > 1$ is arbitrarily large (with or without the transversality condition) - a result proved for $v$ analytic (see \cite{BG}, \cite{B}).

A third problem is to improve the modulus of continuity (\ref{modconi}) of the Lyapunov exponent and of the IDS to say, H\"{o}lder continuity, as in the analytic case.

Finally, of course, the membership to a Gevrey class should not be the ultimate regularity condition on a function $v$ defining the potential of the operator (\ref{op1}). One should consider more general Carleman-class functions, or even functions in a Sobolev space, and see if similar localization results can be obtained.
In Remark \ref{whereitfails1} and Remark \ref{whereitfails2}, we mentioned the obstacles we had in using our method for these more general problems.
}
\end{remark}

\newpage

\section*{Acknowledgments}

The author is grateful to Christoph Thiele, his thesis adviser, for his help and guidance throughout this work.

The author would also like to thank Jean Bourgain (for suggesting the problem, for pointing out the paper \cite{E} and for a useful discussion) to Svetlana Jitomirskaya (for reading the draft and for making useful comments on it) and to Wilhelm Schlag (for suggesting the question of regularity of the Lyapunov exponent and of the IDS in this context, for reading the draft and for a useful discussion about this subject).

\nocite{*}
\providecommand{\bysame}{\leavevmode\hbox to3em{\hrulefill}\thinspace}
\providecommand{\MR}{\relax\ifhmode\unskip\space\fi MR }
\providecommand{\MRhref}[2]{%
  \href{http://www.ams.org/mathscinet-getitem?mr=#1}{#2}
}
\providecommand{\href}[2]{#2}

\end{document}